\documentclass[journal=jpcafh,manuscript=article]{achemso}

\usepackage[version=3]{mhchem} 
\usepackage[english]{babel}
\usepackage[utf8x]{inputenc}
\usepackage[T1]{fontenc}
\usepackage{gensymb}
\usepackage{amsmath}
\usepackage{amssymb}
\usepackage{makecell}
\usepackage{subcaption}

\author{Elena Patyukova}
\affiliation[Durham University]
{Chemistry Department, Durham University, Durham, DH1 3LE, UK}
\email{patyukova@gmail.com}
\author{Erte Xi}
\affiliation[Procter\&Gamble]
{Procter\&Gamble, Mason Business Center, 8700 Mason Montgomery Road, Mason, Ohio, United States}
\author{Mark R. Wilson}
\affiliation[Durham University]
{Chemistry Department, Durham University, Durham, DH1 3LE, UK}

\title{Phase behavior of correlated random copolymers}

\abbreviations{}
\keywords{Markov copolymer, correlated copolymer, polyvinylalcohol, phase behavior, phase separation}

\begin{document}




\begin{abstract}
In this work, we calculate Flory-Huggins phase diagrams for correlated random copolymers. We achieve it in two steps. At first, we derive a distribution function of two-letter A, B copolymer chains depending on the fraction of A-segments and AB-duplets. Then we use the method of moments, which was developed by Sollich and Cates [Sollich, P.; Cates, M. E.;  Phys. Rev. Lett. 1998, 80, 1365–1368] for polydisperse systems, to reduce the number of degrees of freedom of the computational problem and calculate phase diagrams. We explore how the location of transition points and composition of coexisting phases depends on fractions of A-segments and AB-duplets in a sequence and the degree of polymerization. The proposed approach allows taking into account fractionation, which was shown to affect the appearance of phase diagrams of statistical copolymers.

\end{abstract}

\section{Introduction}
Random copolymers, or statistical copolymers, are polymers composed of at least two monomer units connected in a more or less random manner. Random copolymers are ubiquitous in industrial applications. For example, polyvinyl alcohol (PVA) is used for the creation of water-soluble pouches and capsules. PVA is obtained by post polymerization modification from polyvinyl acetate (PVAc), in which the acetate groups can be either, fully or partially transformed into alcohol groups\cite{Tubbs1966SequencePolyvinyl-acetate, Moritani197713C-Copolymers, Denisova2012ChainCopolymers,Ilyin2014EffectBulk}. 
Pure PVA films with a high degree of hydrolysis are brittle and difficult to dissolve because of a considerable degree of crystallinity\cite{Denisova2013ThermalCopolymers}. Lowering the degree of hydrolysis or adding plasticizers leads to more flexible and soluble PVA films. To understand the conditions of stability of these films with respect to segregation and their physical properties it is important to take into account a copolymer nature of PVA\cite{Squillace2020InfluenceActivity,Briddick2018BloomingFilms}.

Other examples of statistical copolymers include carboxymethyl cellulose (CMC)\cite{Ergun2015Cellulose}, acrylonitrile butadiene styrene (ABS), styrene-butadiene rubber (SBR), membrane electolyte polymers like Nafion\cite{Karimi2019RecentReview}, polyurethanes\cite{Teixeira2007DemixingReactions} and many other commercial materials\cite{Daniele2017Single-phase14-cis-polyisoprene, Gringolts2019OlefinSynthesis}.
The ability to predict polymer material properties based on their chemistry, chain structure, and preparation method is crucial for product design and other industrial applications. 
The molecular structure of statistical AB copolymers is most commonly characterized by the composition (i.e. the fraction of A units). Composition totally describes an ensemble of sequences in the case when there are no correlations between the type of segments at different positions along the chain. In this case, copolymers are called \textit{random}. If there is a correlation between the appearance of different types of segments at different positions along a chain then statistical copolymers are called \textit{correlated}. Correlated copolymers can be described macroscopically in terms of concentrations of duplets (AA, BB, AB, BA), triplets (AAA, AAB, etc.), etc. The more concentrations of n-tuplets are needed to fully describe the system, the less randomness there is in a copolymer sequence. Here we consider the situation when only the fractions of A-segment and AB-duplets (which is equivalent to a fixed concentration of all duplets) are enough to fully characterize the copolymer. If the concentration of AB-duplets in a copolymer sequence is reduced, compared to a truly random copolymer, then such correlated copolymer is referred to as a \textit{blocky} copolymer, meaning that segments of one type tend to be arranged in \textit{blocks}. In the opposite case when a copolymer sequence is enriched in AB-duplets, the copolymer is called \textit{alternating}. 

In the case of PVA, the degree of the blockiness of PVA depends on synthetic route \cite{Moritani197713C-Copolymers,Denisova2012ChainCopolymers}. Saponification of PVA leads to blocky copolymers, while acetylation of previously hydrolyzed PVA produces more random sequences. The appearance of correlations during saponification is due to a larger reaction constant for hydrolyzing VAc segments whose neighbors have already been hydrolyzed.  \cite{Noah1974ThePolymers,Denisova2012ChainCopolymers}.

Blocky or alternating copolymers are also common results of sequential polymerisation \cite{Kim2020TuningIsoprene}. In this case blockiness arises when $k_{\rm{AA}}>>k_{\rm{AB}}$ and $k_{\rm{BB}}>>k_{\rm{BA}}$, where $k_{\rm{IJ}}$ is the reaction rate coefficient describing addition of I-segment to the growing chain with segment J at the end. 

Both models belong to the class of first-order Markov models and with the assumption that copolymerization is stationary (concentrations of monomers are kept fixed), both produce the same types of sequences. 

The equilibrium phase behavior of statistical copolymers is rich and is not yet fully understood. Early works on the phase behavior of random copolymers\cite{Scott1952ThermodynamicsCopolymers, Bauer1985EquilibriumCopolymers,Nesarikar1993PhaseCopolymers} concentrated on considering Flory-Huggins mixtures or copolymers with different compositions (fractions of A-segments in AB-copolymer). As the Flory-Huggins parameter, $\chi$ describing interactions between dissimilar segments, increased, an initially homogeneous mixture of copolymer chains separated into two phases with a different fraction of A-segments. Location of a spinodal point, with respect to separation into two phases, was predicted by Scott: \cite{Scott1952ThermodynamicsCopolymers} $\chi_s=\frac{1}{\rho_2-\rho_1^2}$ where $\rho_2$ and $\rho_1$ were correspondingly the second and the first moment of the distribution with respect to composition. Nesarikar \textit{et al.}\cite{Nesarikar1993PhaseCopolymers} went further and calculated phase diagrams with cloud points and fractions of A-segments in coexisting phases for short Bernoulli copolymers. They showed that as the Flory-Higgins parameter, $\chi$, increased, separation into two, three, four, etc. phases occurred. Importantly, it was also shown that distributions with respect to the fraction of A-segments in coexisting phases, in general, had different shapes, in other words, fractionation took place.   
However, these predictions and calculations were made only for polymers with a small number of segments, $N\approx10-30$, and no correlations along the sequence. 

Another approach to predict phase behavior of random copolymers was proposed by Shakhnovich and Gutin\cite{Shakhnovich1989FormationHeteropolymer} and later developed by other authors \cite{Fredrickson1992MulticriticalMelts,Dobrynin1991FluctuationSystems,Angerman1996MicrophaseCopolymers,Vanderwoude2017EffectsCopolymers,Govorun2017MicrophaseCopolymers,Panyukov1996TheHeteropolymers,Subbotin2002PhaseCopolymers}. It was based on Landau expansion of free energy of a copolymer melt in terms of an order parameter representing a deviation of a local composition from a global composition. Coefficients of expansion were calculated within a mean-field approximation and expressed through single-chain correlation functions averaged over all sequences. This approach was applied to correlated random copolymers by Fredrickson \textit{et al.}\cite{Fredrickson1992MulticriticalMelts}.
It was shown that initially homogeneous melt of blocky copolymers separated first into two macro phases upon an increase in the Flory-Huggins parameter. This transition was closely followed by remixing and forming one microphase separated phase with no long-range order. The period of the microphase had a strong dependence on the temperature and decreased as $L\sim\left(T_s-T\right)^{-1/2}$ as temperature decreased (Flory-Huggins parameter increased). For alternating copolymers, a critical value of sequence correlation $\lambda_C$ was found, such that for $\lambda<\lambda_C$ a direct transition from the homogeneous state into microphase separated state was predicted, without macrophase separation\cite{Fredrickson1992MulticriticalMelts}.

Nesarikar \textit{et al.} \cite{Nesarikar1993PhaseCopolymers} pointed out that the one feature of the Shakhnovich-Fredrickson approach was that it did not take fractionation into account. It implicitly assumed that the distribution with respect to compositions (fraction of A-segments) had the same shape and only the mean value of composition had been changed. This assumption was too strong for non-symmetric comopolymers. The further prediction was that for sufficiently short copolymers, $N\lesssim 60$, the transition to microphase should be preceded by the coexistence of three macro phases.

Recently this discussion was continued by von der Heydt \textit{et. al.}\cite{VonDerHeydt2011Three-phaseCopolymers} who considered a mixture of triblock copolymers, AAA, BBB, ABB, AAB, BAB, ABA, with overall composition $f=0.5$ and varied volume fractions of different sequences to imitate Markovian sequence correlations. They showed that as the Flory-Huggins parameter, $\chi$, increased coexistence of two A- and B-rich macro phases was followed by the coexistence of three phases one of which was the lamellar microphase. Microphase emerged as a shadow phase and was enriched in alternating sequences. So, both microphase separation and three-phase coexistence took place simultaneously.

The latest results show that it is important to take fractionation into account to make a correct prediction of phase behavior\cite{Heydt2010SequenceCopolymers}. In this work, we aim at taking into account fractionation in the framework of the Flory-Huggins theory of blocky copolymers with realistic chain lengths. Therefore, we consider only the possibility of macrophase separation despite knowing about the existence and importance of microphase separation in statistical copolymers. This is done to get a solid reference point for a more refined picture including microphase separation, which may be developed in the future.

In this work, we use the method of moments proposed by Sollich \textit{et al.}\cite{Sollich2007MomentSystems} for polydisperse systems. This method can effectively reduce the number of degrees of freedom of the polydisperse system, otherwise, the system consists of, an order of magnitude, $2^N$ different components and direct solution of the phase equilibrium equations is not possible. To use the method of moments we derive the probability distribution function of copolymer chains with respect to fractions of A-segments and AB-duplets. Then we obtain Flory-Huggins phase diagrams of blocky copolymers and study the dependence of phase diagrams on the fraction of A-segments of the copolymer, chain length, and degree of correlations along the sequence (concentration of AB-duplets). At the end we make a comparison of our results with the work of Nesarikar\textit{et  al.}\cite{Nesarikar1993PhaseCopolymers} and Fredrickson \textit{et al.}\cite{Fredrickson1992MulticriticalMelts}.

The paper is organized as follows. First, we derive a distribution function for correlated copolymers.  Then we use this distribution to apply the moments method and obtain phase diagrams, volumes of coexisting phases, and their density distributions. We finish with a discussion of the results. 

\section{Derivation of a distribution function for Markov copolymers}
To derive a distribution function for Markov copolymers of the first order let's first look at AB random binomial copolymers with chain length $N$, the number of A-monomers equal to $N_{\rm{A}}$ and their average fraction $f=\langle N_{\rm{A}}/N\rangle$ in the population of copolymer chains. The distribution function can be written as:
\begin{equation}
    \rho\left(N_{\rm{A}}\right)=\frac{N!}{N_{\rm{A}}!\left(N-N_{\rm{A}}\right)!}f^{N_{\rm{A}}}\left(1-f\right)^{N-N_{\rm{A}}}
\end{equation}.
Or if we use Stirling's formula $N!\approx \sqrt{2\pi N}\left(\frac{N}{e}\right)^N$ and introduce $\sigma=\frac{N_{\rm{A}}}{N}$,
\begin{equation}
\begin{split}
    \rho\left(\sigma\right)=&\frac{1}{\sqrt{2\pi N\sigma\left(1-\sigma\right)}}\left(\frac{f}{\sigma}\right)^{N\sigma}\left(\frac{1-f}{1-\sigma}\right)^{N-N\sigma}\sim \\ &\sim\frac{1}{\sqrt{2\pi 
    Nf\left(1-f\right)}}\left(\frac{f}{\sigma}\right)^{N\sigma}\left(\frac{1-f}{1-\sigma}\right)^{N-N\sigma}
    \end{split}
\end{equation}
Let us consider now an infinite random AB sequence characterized by the fraction of A-segments, $f$. Then we can write down an information rate that corresponds to entropy per monomer of such sequenc\cite{Goldie1991CommunicationTheory, Touchette2009TheMechanics}:
\begin{equation}
    S_{\rm{inf}}=-f\ln{f}-\left(1-f\right)\ln\left(1-f\right)
    \label{entropy-ran}
\end{equation}
The \textit{information chemical potential} of species A is then 
\begin{equation}
    \mu_{\rm{inf}}=-\frac{\partial S_{\rm{inf}}}{\partial f}=\ln{f}-\ln{\left(1-f\right)}
\end{equation}
If there is a finite sequence of length $N$ in equilibrium with this infinite system then its grand potential depending on the number $N_{\rm{A}}=\sigma N$ of A-segments in this sequence is
\begin{equation}
    \frac{\Phi\left(\sigma\right)}{k_{\rm{B}}T}=-NS_{\rm{inf}}\left(\sigma\right)-\mu \left(f\right) N\sigma
\end{equation}
And the probability to have $N_{\rm{A}}=\sigma N$ segments is
\begin{equation}
    \rho\left(\sigma\right)=\frac{1}{Z}e^{-\frac{\Phi\left(\sigma\right)}{k_{\rm{B}}T}}=\frac{1}{Z}\left(\frac{f}{\sigma}\right)^{N\sigma}\left(\frac{1-f}{1-\sigma}\right)^{N-N\sigma}
\end{equation}
Which is a binomial distribution. 

Now let us turn to the first-order Markov sequences. Information rate, in this case, is known to take the form\cite{Goldie1991CommunicationTheory}:
\begin{equation}
    S_{\rm{inf}}=-n_{\rm{AA}}\ln \frac{n_{\rm{AA}}}{p_{\rm{A}}}-n_{\rm{AB}}\ln \frac{n_{\rm{AB}}}{p_{\rm{A}}}-n_{\rm{BA}}\ln \frac{n_{\rm{BA}}}{p_{\rm{B}}}-n_{\rm{BB}}\ln \frac{n_{\rm{BB}}}{p_{\rm{B}}}
\end{equation}
here $p_{\rm{A}}=f$ is the probability of a randomly chosen segment to be type $A$, $p_{\rm{B}}=1-f$ is the probability of a randomly chosen segment to be type $B$, and $n_{\rm{IJ}}$ is a concentration of  IJ-duplets in the sequence. Concentrations of duplets satisfy conditions: $n_{\rm{AA}}+n_{\rm{AB}}+n_{\rm{BA}}+n_{\rm{BB}}=1$,  $n_{\rm{AB}}=n_{\rm{BA}}=\theta$, $n_{\rm{AA}}+n_{\rm{BA}}=f$ and $n_{\rm{BA}}+n_{\rm{BB}}=1-f$, where $f$ is the fraction of A units and $\theta$ is the concentration of AB-pairs. $\theta$ was referred previously as a \textit{block character} by Moritani and Fujiwara\cite{Moritani197713C-Copolymers} as it controls average lengths of blocks when composition of copolymer is fixed, $\langle l\rangle=\frac{f}{\theta}$. In completely random copolymer $\theta = f\left(1-f\right)$ and the average length of the A-block is $\langle l\rangle=\frac{f}{\theta}=\frac{1}{1-f}$. If $\theta < f\left(1-f\right)$ then the sequence is depleted in AB and BA duplets and the copolymer is defined as \textit{blocky}, the average lengths of A-blocks and B-blocks are larger than in random copolymer. Conversely, if $\theta > f\left(1-f\right)$ the concentration of AB and BA duplets is increased and the copolymer tends to be alternating, the average length of A and B-blocks is decreased. The maximum value the $\theta$ can take is $\rm{min}(f,1-f)$. 

An alternative parameter, which is often used to characterise the degree of correlations and 
the average block length in first-order Markov copolymers, is a parameter $\lambda$, introduced by Fredrickson \textit{et al.}\cite{Fredrickson1992MulticriticalMelts}. $\lambda$ describes the degree of correlations of the sequence.
For example, if, at a position $s$ along the chain, there is a segment of type I, then the conditional probability that "at position $s+l$ there is a segment of type J" is proportional to $\lambda^l$.  $\lambda=0$ corresponds to a random copolymer, $\lambda<0$ corresponds to an alternating copolymer and $\lambda>0$ corresponds to a blocky copolymer. Mathematically, $\lambda$ is the non-trivial eigenvalue of the transfer matrix (the trivial eigenvalue equals one), which in our notations takes the form:
\begin{equation}
    \begin{pmatrix}
    p_{\rm{AA}} & p_{\rm{BA}} \\
    p_{\rm{AB}} & p_{\rm{BB}}
    \end{pmatrix}=
    \begin{pmatrix}
    \frac{f-\theta}{f} & \frac{\theta}{1-f} \\
    \frac{\theta}{f} & \frac{1-f-\theta}{1-f}
    \end{pmatrix}
    \label{matrix}
\end{equation}
Therefore, we get for $\lambda$ expressed in terms of $f$ and $\theta$:
\begin{equation}
    \lambda=\frac{f-f^2-\theta}{f\left(1-f\right)}.
\end{equation}
Returning to the information rate and applying conditions on $n_{\rm{IJ}}$ we get:
\begin{equation}
\begin{split}
    S_{\rm{inf}}=&-\left(f-\theta\right)\ln \left(f-\theta\right)-2\theta\ln \theta-\left(1-f-\theta\right)\ln \left(1-f-\theta\right)+\\
    +&f\ln f+\left(1-f\right)\ln\left(1-f\right).
    \end{split}
\end{equation}
This expression depends on two parameters $f$ and $\theta$ which describe the concentration of A-segments and AB-pairs in the infinite sequence correspondingly. In analogy with the case of a random sequence, we calculate the chemical potentials of these pseudo-species from the information rate:
\begin{equation}
    \mu_{\rm{f,inf}}=-\frac{\partial S_{\rm{inf}}}{\partial f}=-\ln f+\ln \left(1-f\right)+\ln \left(f-\theta\right)-\ln \left(1-f-\theta\right)
\end{equation}
\begin{equation}
    \mu_{\rm{\theta,inf}}=-\frac{\partial S_{\rm{inf}}}{\partial \theta}=2\ln \theta-\ln \left(1-f-\theta\right)-\ln \left(f-\theta\right)
\end{equation}
For a distribution function for a chain with the finite length $N$, the fraction of A-segments $\sigma$ and the fraction of AB-pairs $t$ which is in equilibrium with an infinite chain with the composition $f$ and the fraction of AB-duplets $\theta$, we get:
\begin{equation}
\begin{split}
    \rho\left(\sigma,t\right)=&\frac{1}{Z}e^{NS_{\rm{inf}}\left(\sigma,t\right)+\mu_{\rm{f,inf}} N\sigma+\mu_{\rm{\theta,int}}Nt}=\\
    =&\frac{1}{Z}\left(\frac{f}{\sigma}\right)^{-N\sigma}\left(\frac{1-f}{1-\sigma}\right)^{-N+N\sigma}\left(\frac{f-\theta}{\sigma-t}\right)^{N\sigma-Nt}\left(\frac{\theta}{t}\right)^{2Nt}\left(\frac{1-f-\theta}{1-\sigma-t}\right)^{N-N\sigma-Nt}
    \end{split}
\end{equation}
Where $Z$ is determined from the normalization condition
\begin{equation}
   \int_0^1 d\sigma\int_0^{\rm{min}\left(\sigma,1-\sigma\right)} \rho\left(\sigma,t\right) dt =1
\end{equation}

If we reverse the Stirling approximation again in analogy with a binomial distribution, we can get:
\begin{equation}
\begin{split}
    \rho\left(\sigma,t\right)=&\frac{\left(N\sigma\right)!\left(N-N\sigma\right)!}{\left(N\sigma-Nt\right)!\left(Nt\right)!^2\left(N-N\sigma-Nt\right)!}\\
    &\frac{\left(f-\theta\right)^{N\sigma-Nt}\theta^{2Nt}\left(1-f-\theta\right)^{N-N\sigma-Nt}}{f^{N\sigma}\left(1-f\right)^{N-N\sigma}}
    \end{split}
    \label{distribution}
\end{equation}
In Appendix B we show another way to derive this distribution, which serves as additional support to the calculations presented above.  

We would like to note that similar ideas were developed previously to characterize the local composition profile of compatible polymer blends in the course of macromolecular reactions and interdiffusion\cite{Yashin1997MacromolecularBlend, Kudryavtsev2006Diffusion-inducedCopolymers}. 

\begin{figure}[ht!]
    \centering
    \begin{subfigure}[b]{0.45\textwidth}
    \includegraphics[width=\textwidth]{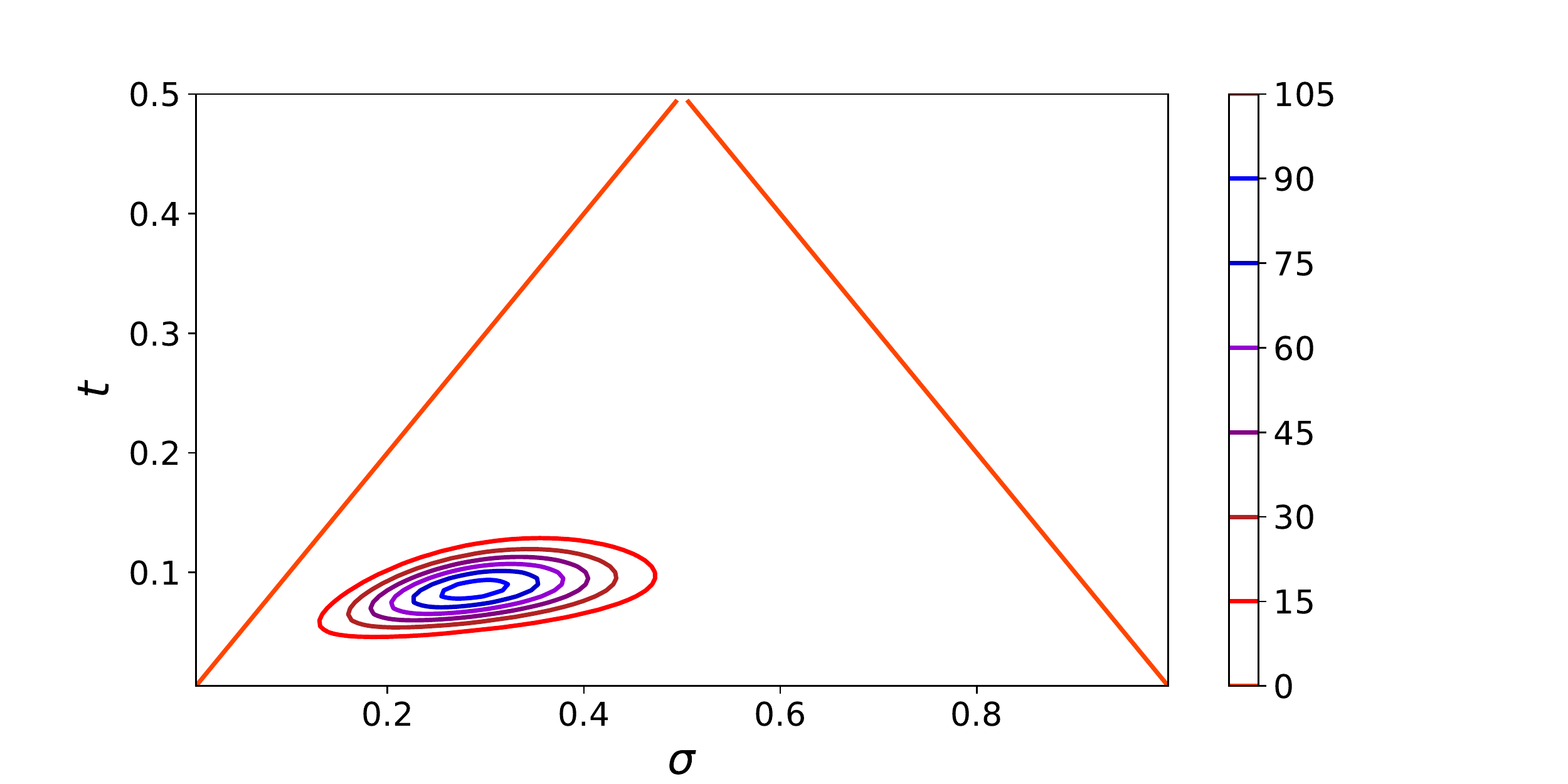}
    \caption{}
    \label{h-sim}
    \end{subfigure}
    \begin{subfigure}[b]{0.45\textwidth}
    \includegraphics[width=\textwidth]{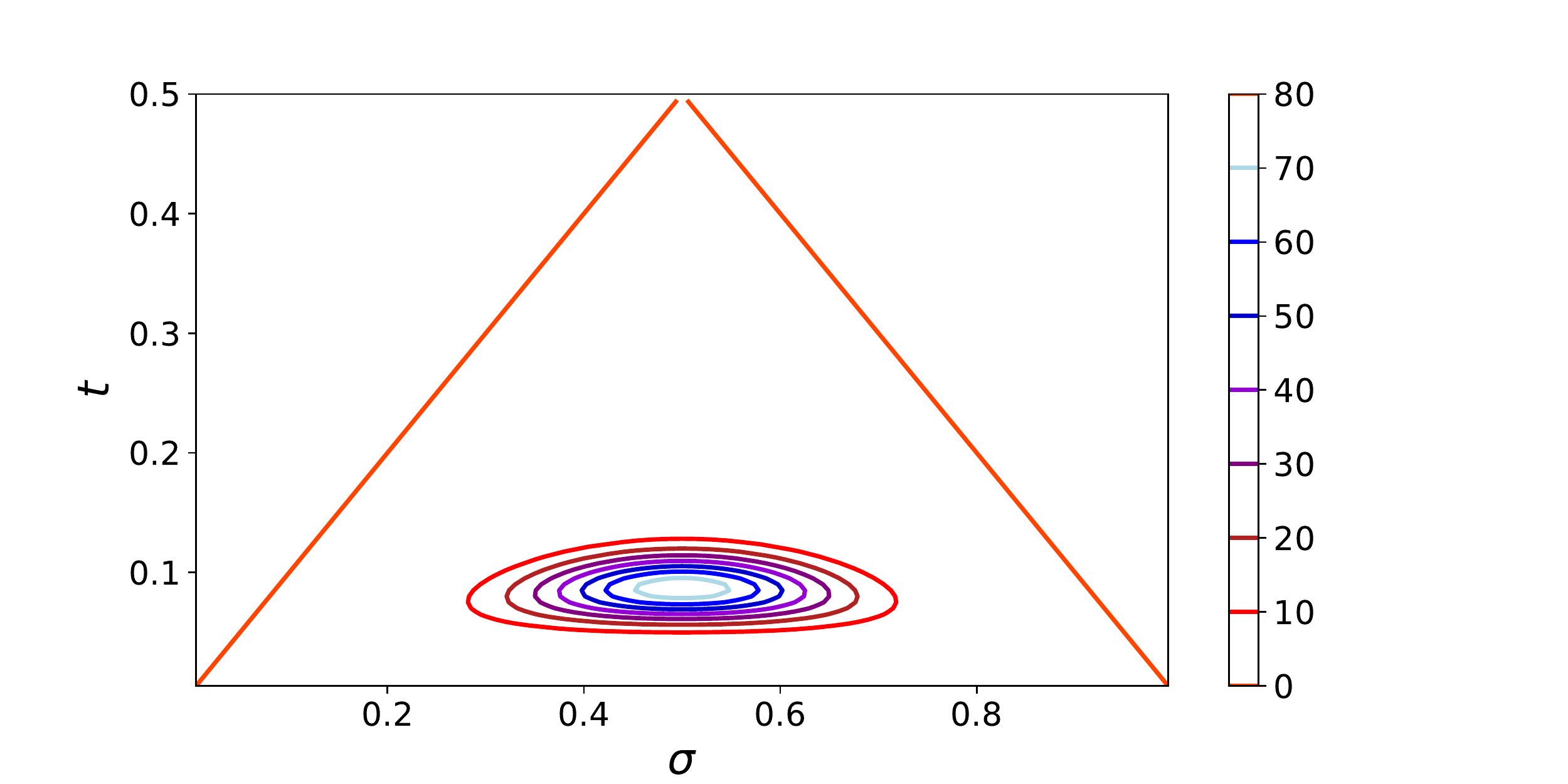}
    \caption{}
    \label{h-theory10}
    \end{subfigure}
    \begin{subfigure}[b]{0.45\textwidth}
    \includegraphics[width=\textwidth]{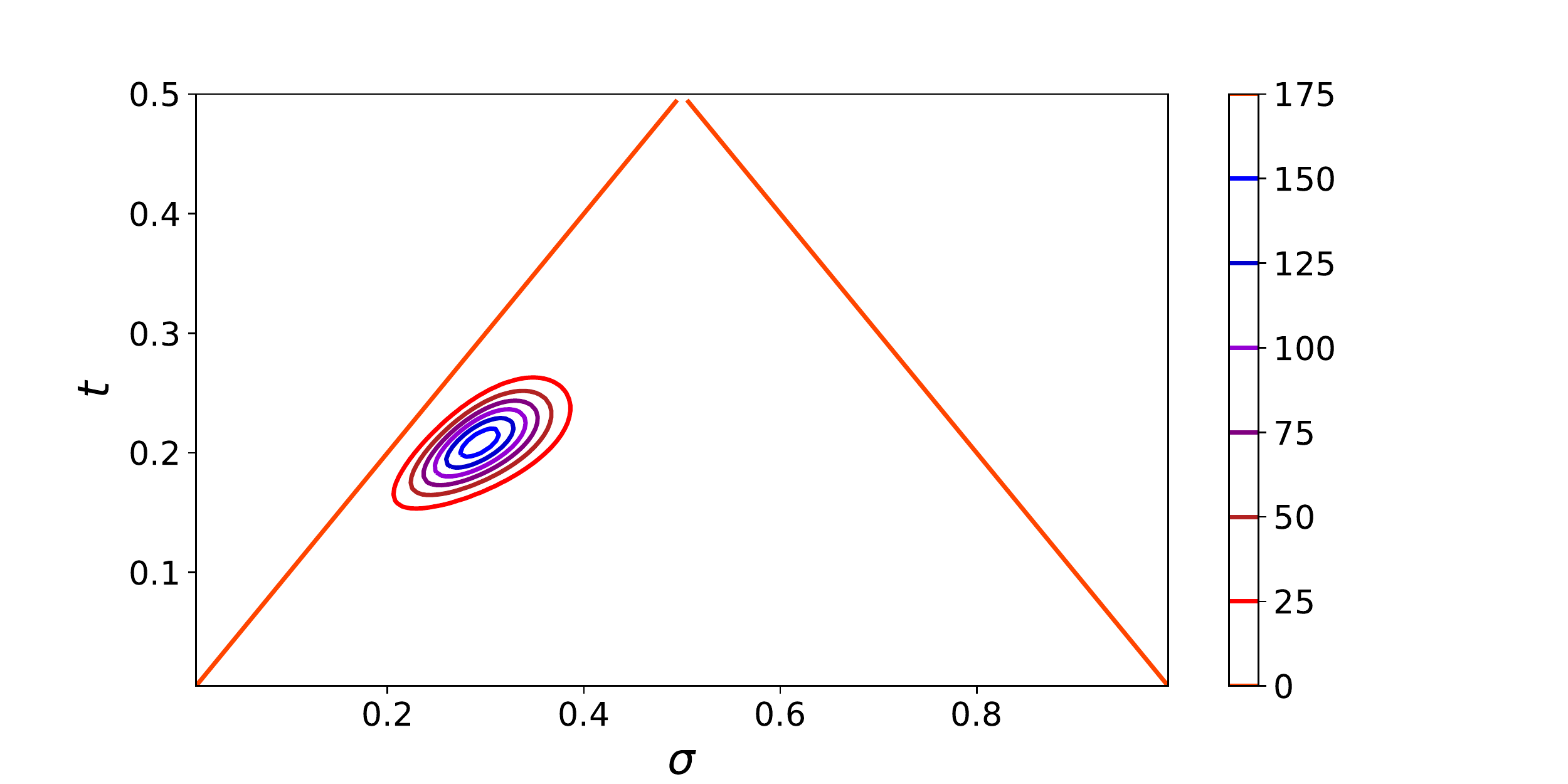}
    \caption{}
    \label{}
    \end{subfigure}
    \begin{subfigure}[b]{0.45\textwidth}
    \includegraphics[width=\textwidth]{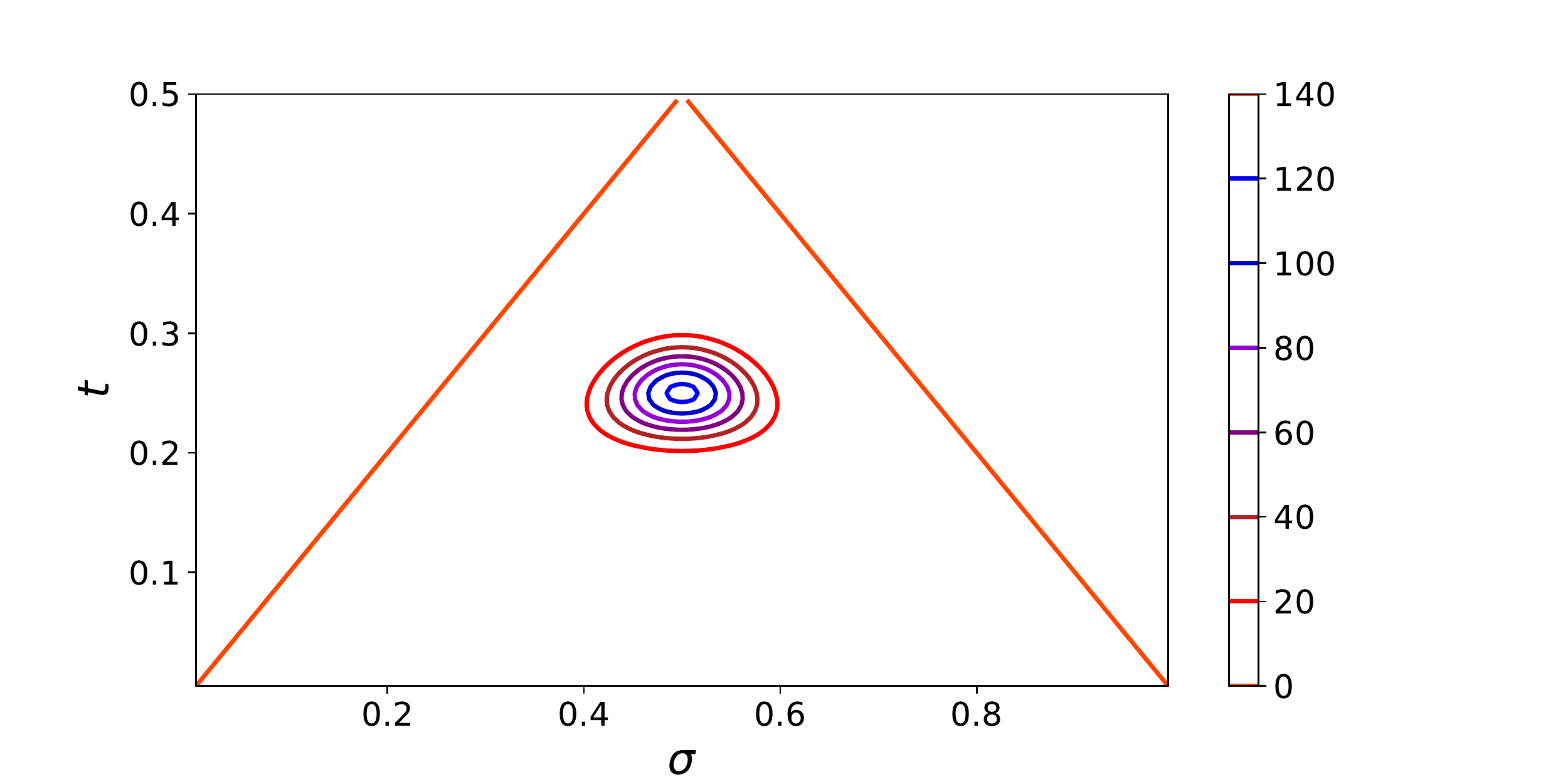}
    \caption{}
    \label{}
    \end{subfigure}
    \begin{subfigure}[b]{0.45\textwidth}
    \includegraphics[width=\textwidth]{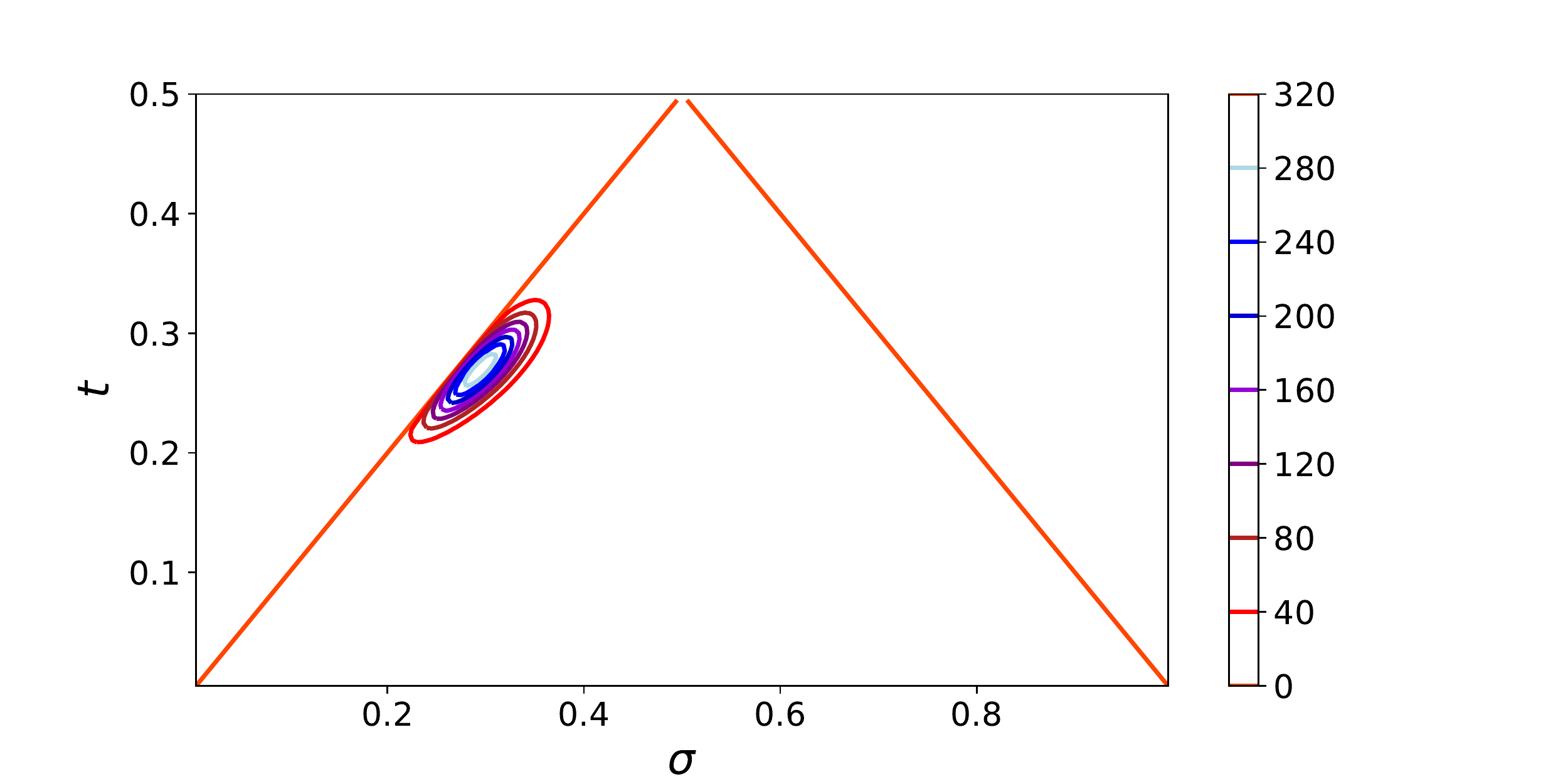}
    \caption{}
    \label{}
    \end{subfigure}
    \begin{subfigure}[b]{0.45\textwidth}
    \includegraphics[width=\textwidth]{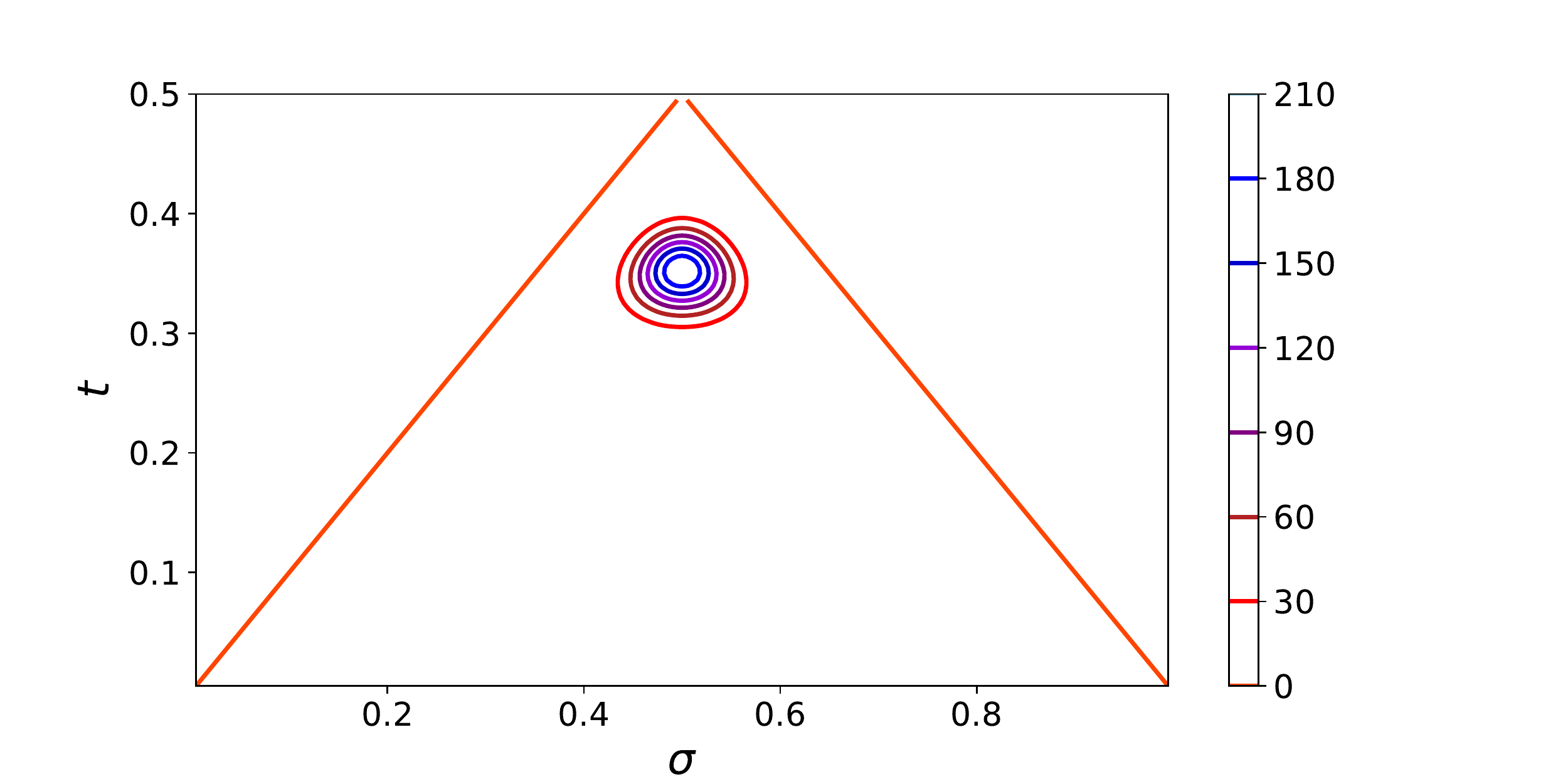}
    \caption{}
    \label{}
    \end{subfigure}
    \caption{Distribution $\rho\left(\sigma,t\right)$ for copolymers characterised by $N=100$ and (a) $f=0.3$,  $\theta=0.09$ (blocky); (b) $f=0.5$,  $\theta=0.09$ (blocky); (c) $f=0.3$, $\theta=0.21$ (random); (d)$f=0.5$, $\theta=0.25$ (random); (e)$f=0.3$,  $\theta=0.27$ (alternating); (f) $f=0.5$, $\theta=0.35$ (alternating).}
    \label{contour}
\end{figure}

Example contour plots of the distributions are shown in Figure \ref{contour}. We can see that at a fixed value of $N$ and $f$, the width of the distribution projected on the $\sigma$ axes decreases as $\theta$ increases. We expected to see this because the difference in compositions between different chains is expected to be larger when segments are arranged in blocks. If we compare two distributions with the same $\theta$ but different $f$ we can see that the distribution with $f$ closer to $0.5$ is broader, so variations in composition are largest for the symmetric copolymer. 

\section{Moment free energy}
Let us now consider a melt of random copolymers with the distribution $\rho\left(\sigma,t\right)$ and write down a moment free energy for it\cite{Sollich2007MomentSystems}. We start with Flory-Huggins free energy for the melt:
\begin{equation}
    \frac{F}{k_{\rm{B}}TV}=\frac{1}{N}\int \rho\left(\sigma,t\right)\left(\ln\frac{\rho\left(\sigma,t\right)}{R_0\left(\sigma,t\right)}-1\right) d\sigma dt +\chi\rho_1\left(1-\rho_1\right)
\end{equation}
where $V$ is the volume of the system, $\chi$ is a Flory-Huggins parameter describing interactions of segments of type A and B and
\begin{equation}
    \rho_1=\int \sigma \rho\left(\sigma,t\right)d\sigma dt
\end{equation}
is the total volume fraction of A-monomers. $R_0\left(\sigma,t\right)$ is the distribution of chains with respect to fractions of A-segments and AB-duplets in the parent phase. It is added here for convenience as far as it represents a term linear in density $\rho\left(\sigma,t\right)$, which does not affect the phase behavior of a copolymer melt. We also implicitly assume that volumes of A and B segments are equal to each other, both segments are flexible so that $\rho\left(\sigma,t\right)$ is simply a volume fraction of the polymer with parameters $\sigma$, $t$.\bibnote{Phase behavior of stiff copolymers in the framework of wormchain model was recently studied by the group of A. J. Spakowitz, for example \cite{Mao2016ImpactCopolymers}}

Now we can apply the moments method\cite{Sollich2007MomentSystems} to reduce the number of degrees of freedom of the problem. 
We fix $m$ first moments of the distribution $\rho_i=\int \sigma^i \rho\left(\sigma,t\right)d\sigma dt$ using Lagrange multipliers $\lambda_i$ along with the total volume fraction which is equal to 1 with Lagrange multiplier $\lambda_0$ and the constrained free energy is minimized with respect to the remaining degrees of freedom: 
\begin{equation}
\begin{split}
     \frac{F'}{kTV}=&\frac{1}{N}\int \rho\left(\sigma,t\right)\left(\ln\frac{\rho\left(\sigma,t\right)}{R_0\left(\sigma,t\right)}-1\right) d\sigma dt +\chi\rho_1\left(1-\rho_1\right)-\\
     -&\lambda_0\left(\int  \rho\left(\sigma,t\right)d\sigma dt - 1 \right)-\sum_{i=1}^{m}\lambda_i\left(\int \sigma^i \rho\left(\sigma,t\right)d\sigma dt - \rho_i \right).
     \end{split}
     \label{initial}
\end{equation}
We note here that though we have 2D-distribution, depending on two parameters, $\rho\left(\sigma,t\right)$, we fix only moments of the composition here. The reason for this is that the Flory-Huggins interaction term depends only on the first moment of the $\sigma$ (fraction of A-segments), so fixing of other moments (for example, moments of $t$) does not affect the phase diagram. This was checked by direct calculations (fixing all moments up to the 3rd order gives the same result as fixing only moments of $\sigma$). In case when the contribution of interactions into free energy depends additionally on $\theta_1=\langle t \rangle$, the average concentration of AB-duplets (physically this case takes place, for example, when Flory-Huggins parameter describing interactions of two segments depends on the type of their neighbors (the model considered by Balazs, \textit{et al.}\cite{Balazs1985EffectBlends}), both moments should be fixed.

Minimizing $F'$ with respect to $\rho\left(\sigma,t\right)$ we get 
\begin{equation}
    \rho\left(\sigma,t\right)=R_0\left(\sigma,t\right)e^{\lambda_0+\sum_{i=1}^{m}\lambda_i\sigma^i}.
    \label{family}
\end{equation}
In this approximation, all possible distribution functions in any coexisting phases belong to this family (equation \ref{family}). It is clear why we needed to include $R_0\left(\sigma,t\right)$ in the formula (equation \ref{initial}), because the parent phase must be included in the family. The larger the number of fixed moments, $m$, the larger the set of functions for approximating distributions in daughter phases. Next, we substitute equation \ref{family} into equation \ref{initial} and obtain an expression for the moment free energy (omitting terms which are linear in $\rho_i$):
\begin{equation}
    F_m=\lambda_0+\sum_{i=1}^m\lambda_i\rho_i-\chi'\rho_1^2,
\label{moment-free}    
\end{equation}
where $\chi'=\chi N$. This expression can be analysed further as the free energy of a m-component system.

\section{Phase diagrams}
In this section we calculate phase diagrams using derived distribution (\ref{distribution}) and moments free energy (\ref{moment-free}). 

We expect that upon an increase of the Flory-Huggins parameter $\chi'$ an initially uniform melt separates into two phases at a cloud point $\chi'_{\rm{cloud}}$, where a new phase with infinitesimal volume emerges. It is also expected that the spinodal point of the system is located at some value of $\chi'_s > \chi'_{\rm{cloud}}$. Beyond the spinodal point, the homogeneous state can't exist as metastable.

To construct a phase diagram and to determine the composition of coexisting phases from the moment free energy we write down standard expressions for chemical potentials and osmotic pressures.
\begin{equation}
\mu_{1}=\frac{\partial F_m}{\partial \rho_1}=\lambda_1-2\chi'\rho_1 
\end{equation}
\begin{equation}
\mu_{i}=\frac{\partial F_m}{\partial \rho_i}=\lambda_i, i>1
\end{equation}
\begin{equation}
\pi=-F_m+\sum_{i=1}^m\rho_i\mu_{i}=-\lambda_0-\chi'\rho_1^2.
\end{equation}
Conditions of phase equilibrium for coexisting phases $\alpha$, $\beta$, $\gamma$, ... are:
\begin{equation}
    \mu_{i}^{\alpha}=\mu_{i}^{\beta}=\mu_{i}^{\gamma}=...,1\le i\le m
\end{equation}
\begin{equation}
    \pi^{\alpha}=\pi^{\beta}=\pi^{\gamma}=...,1\le i\le m
\end{equation}
Additionally, there are conditions that amounts of $\rho_i$ are conserved:
\begin{equation}
    \rho_i^{(0)}= v_{\alpha}\rho_i^{\alpha}+v_{\beta}\rho_i^{\beta}+v_{\gamma}\rho_i^{\gamma}+...
\end{equation}
where $\rho_i^{(0)}$ is the i-th moment of composition in the parent phase and
\begin{equation}
    1=v_{\alpha}+v_{\beta}+v_{\gamma}+...
\end{equation}
sum of all volume fractions of coexisting phases equals to 1. 

These equations can be solved by standard Newton method. For example, for 3 coexisting phases with $m$ moments fixed the set of unknowns is $\{\lambda_1^{\alpha},\lambda_1^{\beta},\lambda_1^{\gamma},\lambda_2,..,\lambda_m,v_{\alpha},v_{\beta}\}$ and there are $m+4$ equations in total (see Appendix A for details). 

To determine the first spinodal and cloud points it is enough to fix only one moment\cite{Sollich2007MomentSystems}.
For the spinodal point at which the initial homogeneous phase loses stability analytic expression can be derived:
\begin{equation}
    \frac{\partial^2 F_m}{\partial \rho_1^2}=\frac{\partial\lambda_1}{\partial\rho_1}-2\chi'_s=\frac{1}{\rho_2-\rho_1^2}-2\chi'_s=0,
    \label{spinodal}
\end{equation}
which is a well-known condition for the spinodal point in random copolymers. \cite{Scott1952ThermodynamicsCopolymers}

To calculate characteristics of coexisting phases above the cloud point it is needed to fix more moments. However, it is not known in advance how many moments are needed to be fixed to produce a satisfactory approximation of the actual composition of the coexisting phases. We proceed by fixing an increasing number of moments at each step until the phase diagram does not change anymore for a given value of $\chi N$.

\begin{figure}[ht!]
    \centering
    \begin{subfigure}[b]{0.8\textwidth}
    \includegraphics[width=\textwidth]{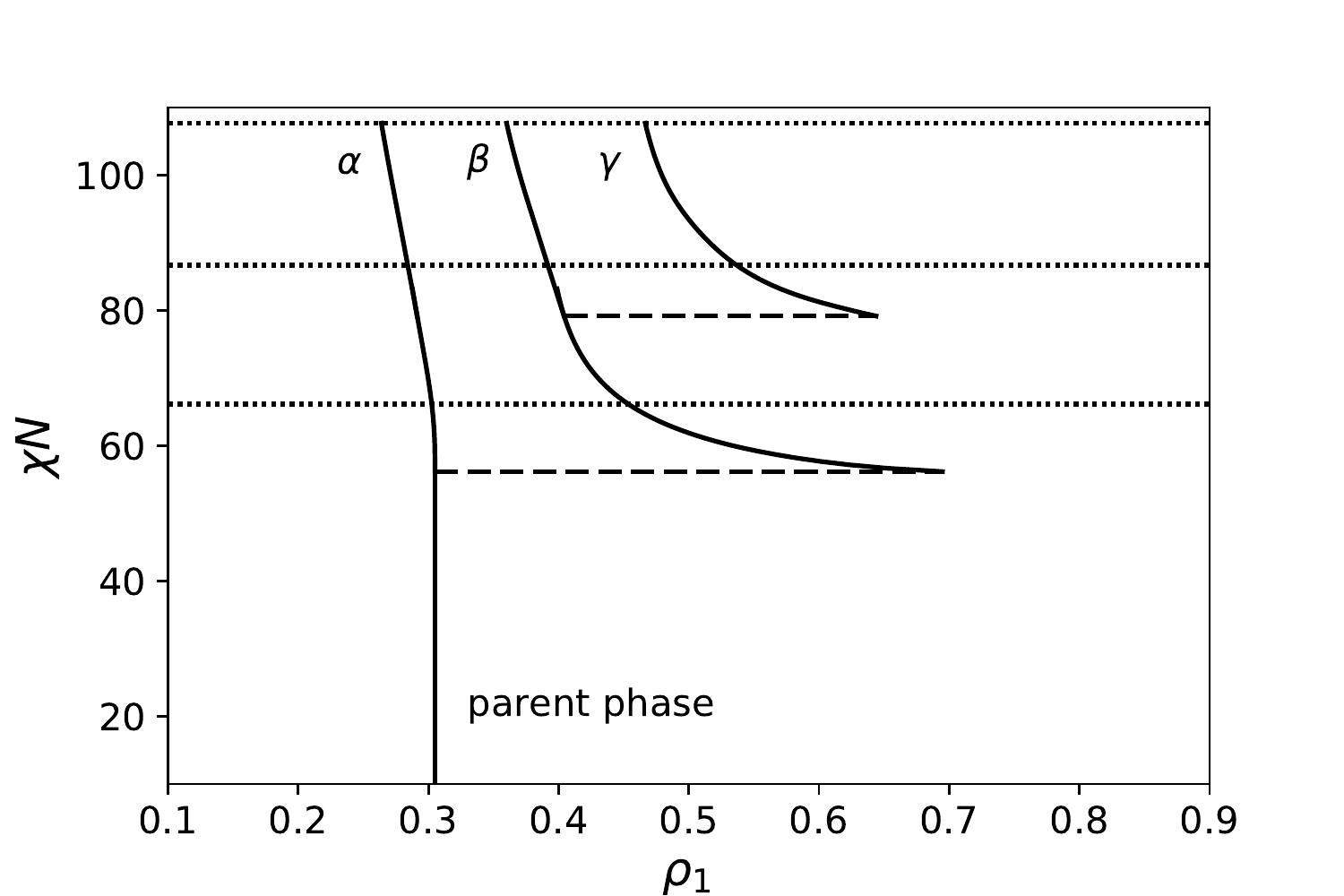}
    \caption{}
    \label{f03t009}
    \end{subfigure}
    \begin{subfigure}[b]{0.8\textwidth}
    \includegraphics[width=\textwidth]{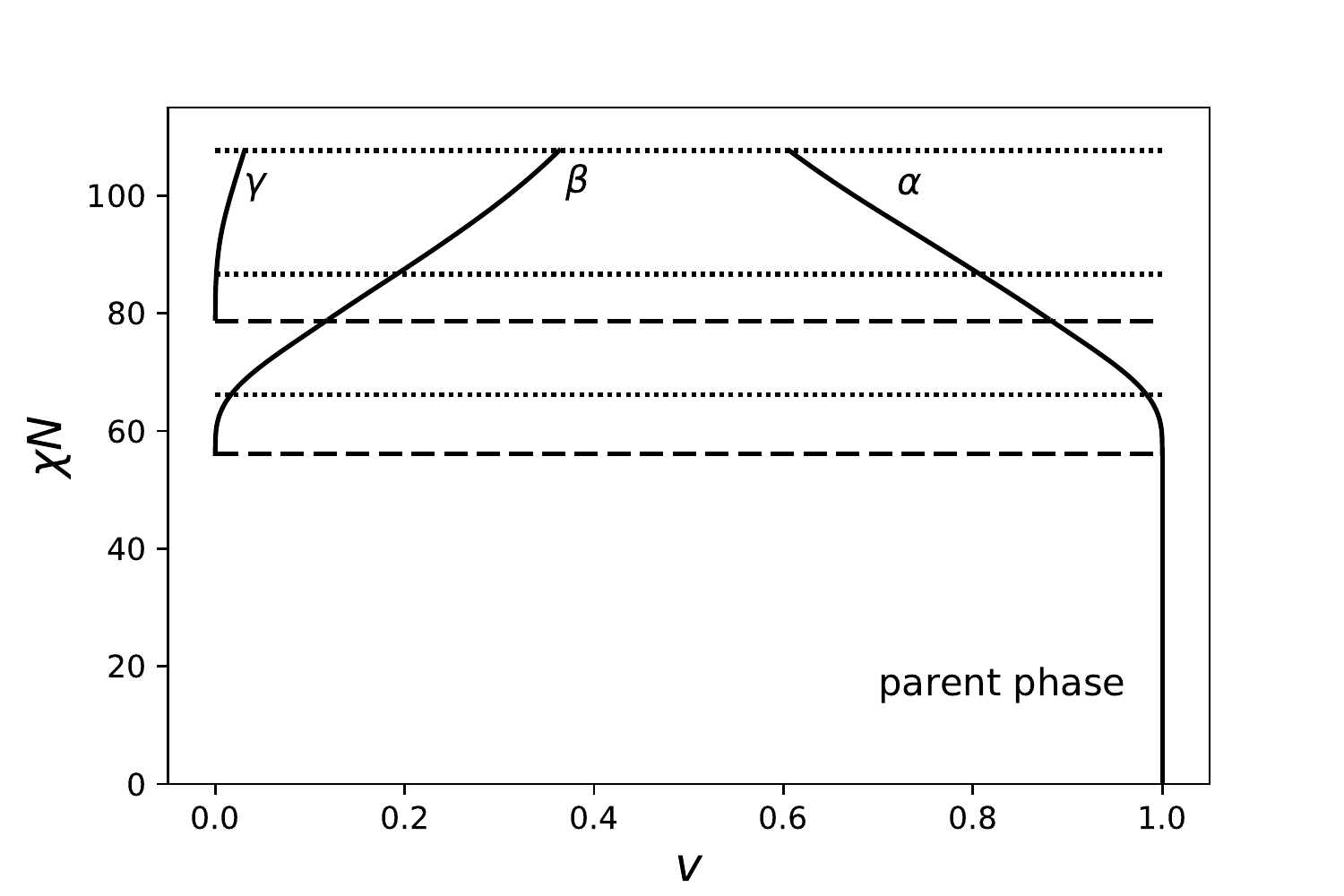}
    \caption{}
    \label{f03t009-volume}
    \end{subfigure}
    \caption{(a) Phase diagram for  $f=0.3$ and $\theta=0.09$, $N=100$ and 9 moments fixed. (b) volume fractions of coexisting phases. Spinodals are shown as dotted lines, cloud points are shown with broken lines, solid curves show compositions of coexisting phases in sub-figure (a) and volume fractions of coexisting phases ($\alpha$, $\beta$, $\gamma$) in sub-figure (b). As Flory-Huggins $\chi$ parameter increases the number of coexisting phases increases.}
\end{figure}

\begin{figure}
    \centering
    \includegraphics[width=0.8\textwidth]{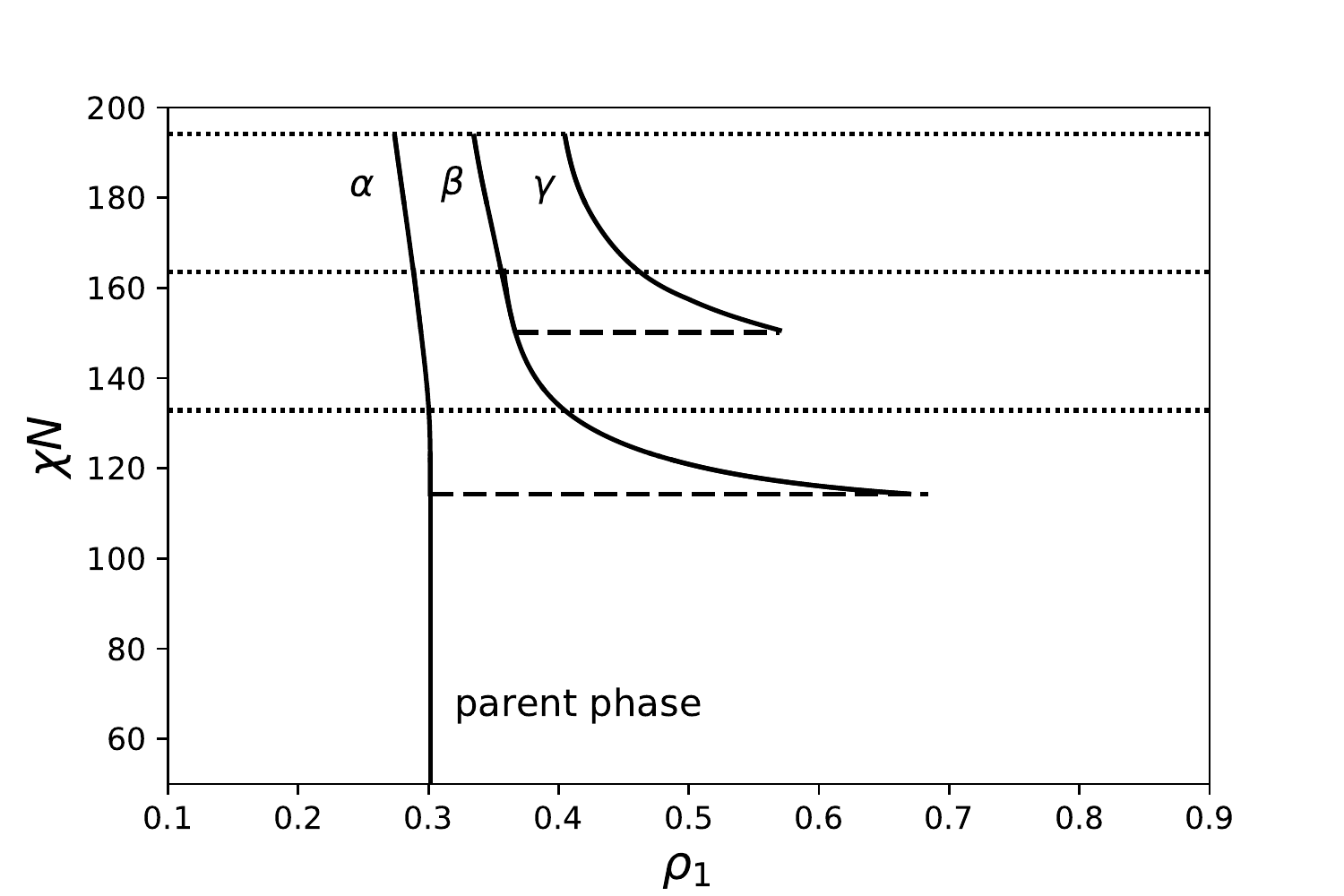}
    \caption{Phase diagrams for $f=0.3$, $\theta=0.15$ and $N=100$ and 9 moments fixed. Spinodals are shown as dotted lines, cloud points are shown with broken lines, solid curves represent compositions of coexisting phases.}
    \label{f03t015}
\end{figure}
\begin{figure}
    \centering
    \includegraphics[width=0.8\textwidth]{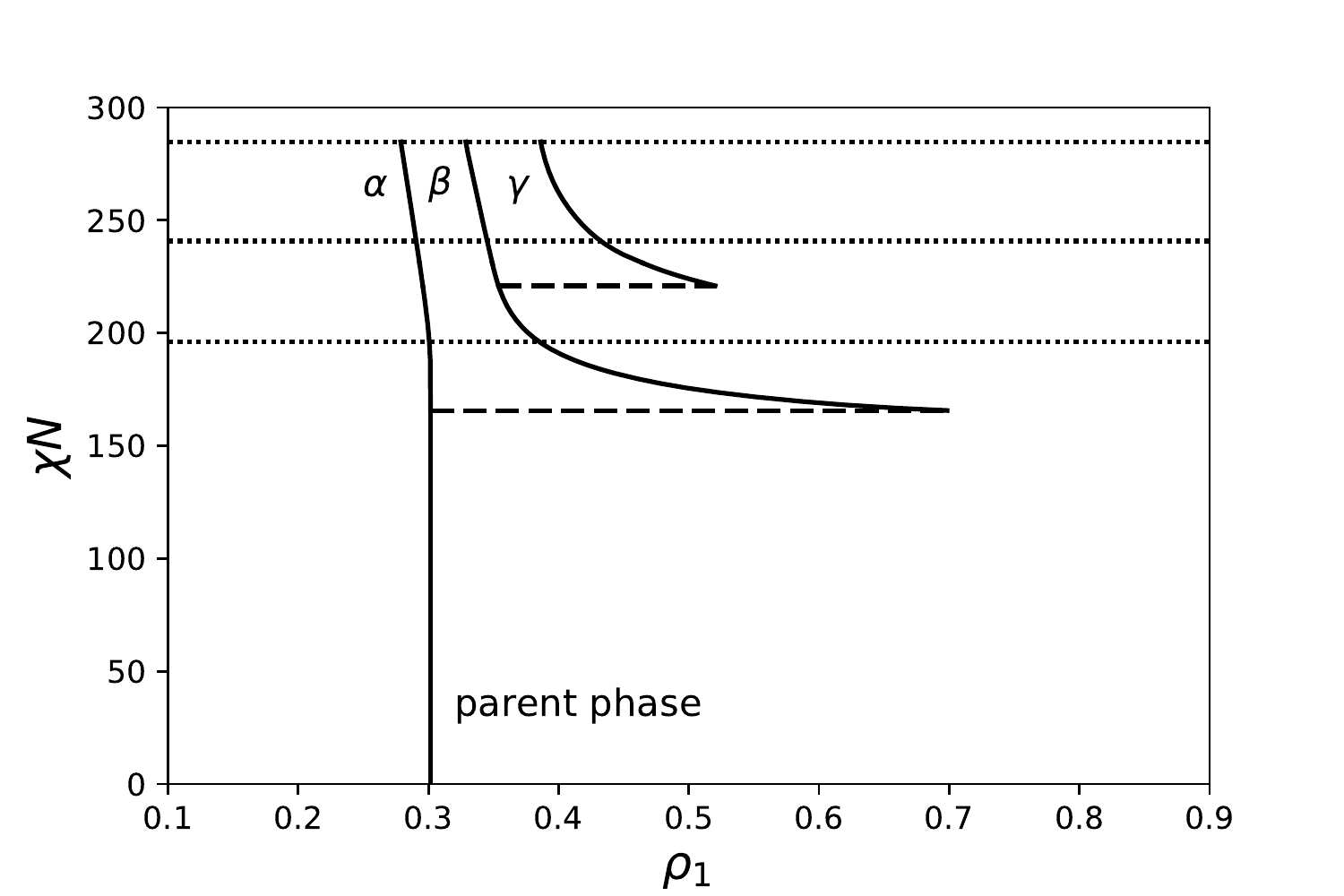}
    \caption{Phase diagrams for $f=0.3$, $\theta=0.09$ and $N=300$ and 9 moments fixed. Spinodals are shown as dotted lines, cloud points are shown with broken lines, solid curves represent compositions of coexisting phases.}
    \label{f03t009n300}
\end{figure}

\begin{figure}[ht!]
    \centering
    \begin{subfigure}[b]{0.8\textwidth}
    \includegraphics[width=\textwidth]{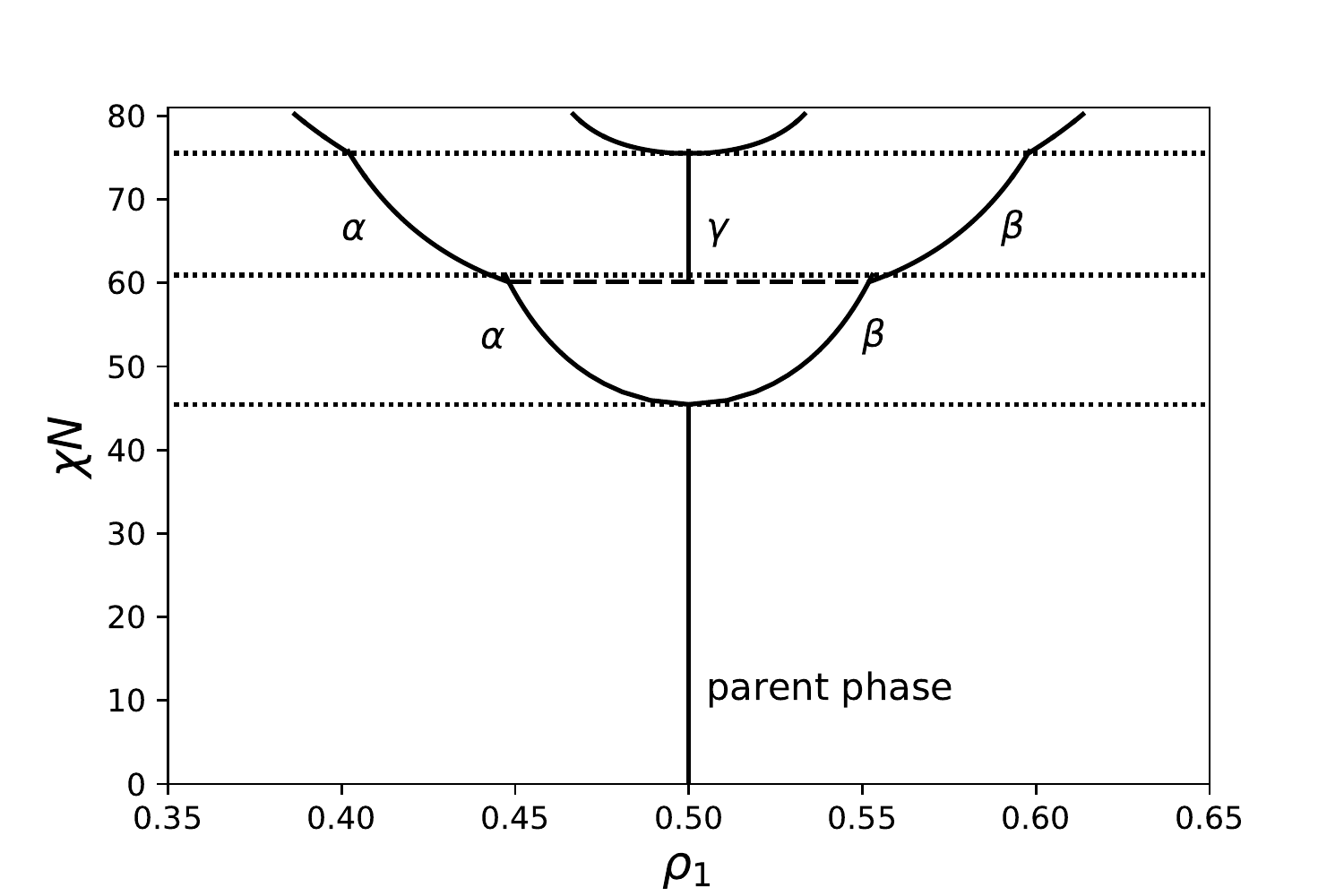}
    \caption{}
    \label{f05t009}
    \end{subfigure}
    \begin{subfigure}[b]{0.8\textwidth}
    \includegraphics[width=\textwidth]{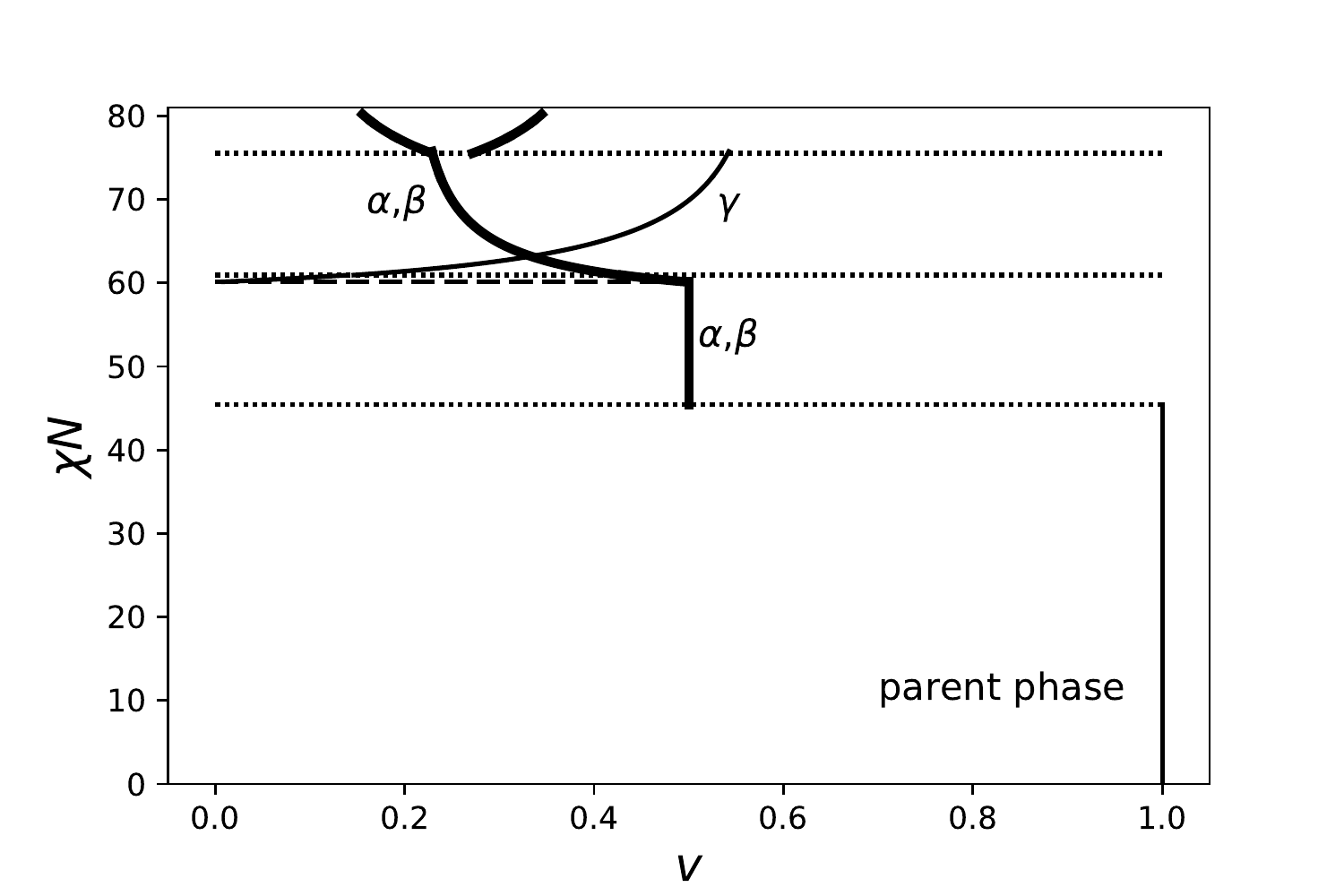}
    \caption{}
    \label{f05t009-volume}
    \end{subfigure}
    \caption{(a) Phase diagrams for $f=0.5$, $\theta=0.09$ and $N=100$.  (b) volume fractions of coexisting phases. Spinodals are shown as dotted lines, cloud points are shown with broken lines, solid curves represent compositions of coexisting phases. }
\end{figure}

Figure \ref{f03t009} shows the phase diagram of a copolymer with fraction of A-units $f=0.3$, fraction of AB-duplets $\theta=0.09$ and chain length $N=100$. Figure \ref{f03t009-volume} shows the dependence of volume fractions of coexisting phases on $\chi N$.  The binodal for separation of initially homogeneous phase into two phases is located at $\chi N=56.2$. One of these phases is the cloud phase with composition equal to the initial composition of the system $\rho_1=0.305$ (it is denoted as $\alpha$). It occupies nearly all the system volume (see Figure \ref{f03t009-volume}). Another phase is a shadow phase with volume fraction of A-segments $\rho_1=0.695$ (it is denoted as $\beta$), it occupies an infinitesimal volume. Upon further increase in $\chi N$, the average fraction of A-segments in the shadow phase decreases strongly until it reaches the value $\rho_1=0.45$ at the spinodal point ($\chi N=66.2$) corresponding to the instability point of the initial homogeneous phase. The decrease in the volume fraction of A-segments in the shadow phase is a consequence of the increase of its volume fraction. If the volume fraction of a phase is sufficiently high its average fraction of A-segments is inevitably close to the volume fraction of A-segments in the initial phase because the initial distribution has one thin single peak. However, it is interesting to note the compositional contrast between phases above the spinodal, for example, at $\chi N=70$ $\Delta \rho_1 \approx 0.13$ is larger than standard deviations of each phase $\left(\rho_2^{\alpha}-\left(\rho_1^{\alpha}\right)^2\right)^{1/2}=0.0829$ and $\left(\rho_2^{\beta}-\left(\rho_1^{\beta}\right)^2\right)^{1/2}=0.0831$ (noting that standard deviation of the parent phase is a bit larger and is equal in this case to $\left(\rho_2^{0}-\left(\rho_1^{0}\right)^2\right)^{1/2}=0.0869$) meaning that they can be distinguished.
At $\chi N\approx77$ the second binodal with respect to the coexistence of three phases is located and a new shadow phase emerges (denoted as $\gamma$). The compositions of coexisting phases at this point are $\rho_1=0.29$, $\rho_1=0.4$ and $\rho_1=0.64$. At $\chi N \approx 86.4$ the first shadow phase losses its stability, which corresponds to the second spinodal point (phase compositions are $\rho_1=0.28$, $\rho_1=0.39$ and $\rho_1=0.54$). Mathematically this means that determinant of the matrix composed of second derivatives of moments free energy with respect to fixed moments $\rho_i$ equals zero at this point in the shadow phase:
\begin{equation}
    \begin{vmatrix}
    \frac{\partial^2 F_m}{\partial \rho_1^2} & ... &\frac{\partial^2 F_m}{\partial \rho_1\partial \rho_m} \\
    ... & ... & ...\\
    \frac{\partial^2 F_m}{\partial \rho_1 \partial \rho_m} & ... &\frac{\partial^2 F_m}{\partial \rho_m^2}
    \end{vmatrix}=0
\end{equation}
The next spinodal point is located at $\chi N = 107.7$ and here the shadow phase with the largest $\rho_1$ loses stability and separates into two phases. However, this fact is not reflected in Figure \ref{f03t009}. Also, the cloud point at which the fourth phase emerges is not shown.

Below all diagrams are calculated with 9 moments of compositions fixed. We discovered that to correctly predict the direction of change in the composition of a shadow phase above the first cloud point (for example, see Figure \ref{f03t009}) at least three moments should be fixed for copolymers with nonsymmetric composition $f\ne0.5$. In order to predict correctly, the location of the second spinodal point and volumes of coexisting phases at least 6 moments need to be fixed. And in order to predict the location of the second cloud point at least 8 moments are required.

Figure \ref{f03t015} shows the phase diagram of a copolymer with composition $f=0.3$, blockiness $\theta=0.15$ and length $N=100$, calculated with moment free energy depending on 9 moments of the distribution. The increased value of $theta$, fraction of AB-duplets, compared to Figure \ref{f03t009} means that copolymer is less correlated. However, this value is still less than in a binomial copolymer with the same composition, i.e. $\theta=f\left(1-f\right)=0.21$. One can see that increase in blockiness leads both to an increase of the value of $\chi N$ at which the first spinodal and the first cloud point are located. It is expected because an increase in blockiness leads to a decrease in the variance of the distribution, so increase of $\chi N$ at spinodal point (equation (\ref{spinodal})). The same reason explains why the distance between consecutive spinodals on the diagram increases and the compositional contrast decreases. Though the composition of a shadow phase at the first cloud point does not change ($f_{sh}=1-f$).


Figure \ref{f03t009n300} shows the effect of increasing $N$ on phase diagram, it is calculated for a copolymer characterised by parameters $f=0.3$, $\theta=0.09$, $N=300$. Interestingly, we can see that as we increase $N$ both the first spinodal point $\chi_S$ and the first cloud point $\chi_C$ slightly decrease and their difference $\Delta \chi = \chi_S - \chi_C$ essentially does not change.  


Finally, Figure \ref{f05t009} shows a special case of a phase diagram for a system with a critical composition $f=0.5$. It differs from the phase diagrams for asymmetric copolymers. The first spinodal point coincides with the cloud point, and the transition from one phase to two phases is continuous. The composition of the two coexisting phases is symmetric with respect to the line $f=0.5$ and their volume fraction does not change until the third phase emerges (Figure \ref{f05t009-volume}). The transition from two phases to three phases is discontinuous. As $\chi N$ increases the volume fraction of a phase with the volume fraction of A-segments $f=0.5$ increases until the next spinodal line. The next transition is again continuous. These observations agree with observations of Nesarikar \textit{et al.} \cite{Nesarikar1993PhaseCopolymers} for binomial copolymers. 

\begin{figure}
    \centering
    \includegraphics[width=0.7\textwidth]{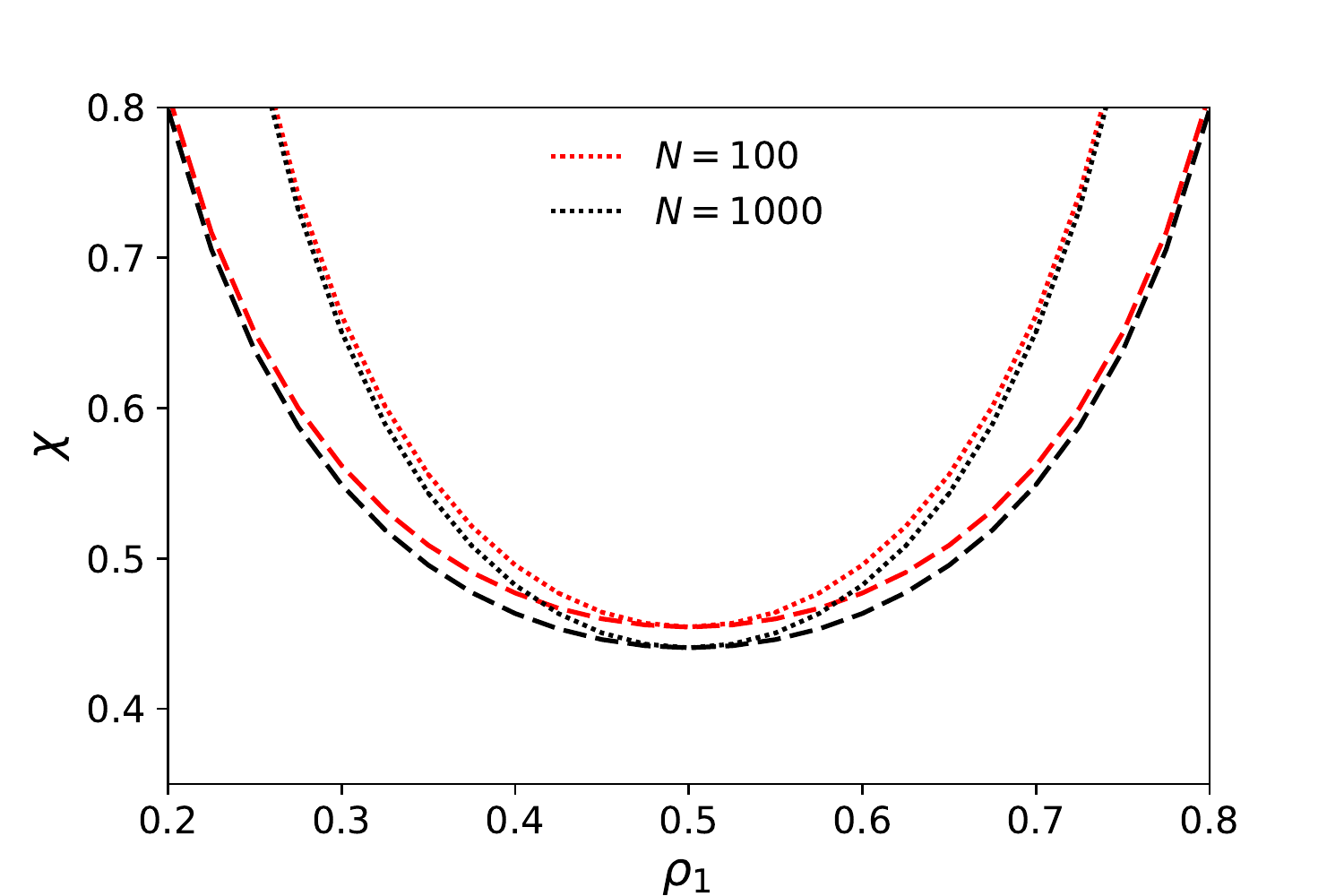}
    \caption{Dependence of spinodal points $\chi_S$ (dotted) and cloud points $\chi_C$(dashed) at fixed values of $\theta=0.09$ and  (1) $N=100$ (red), (2) $N=1000$ (black)}
    \label{spinodal-cloud-t009}
\end{figure}

\begin{figure}
    \centering
    \includegraphics[width=0.7\textwidth]{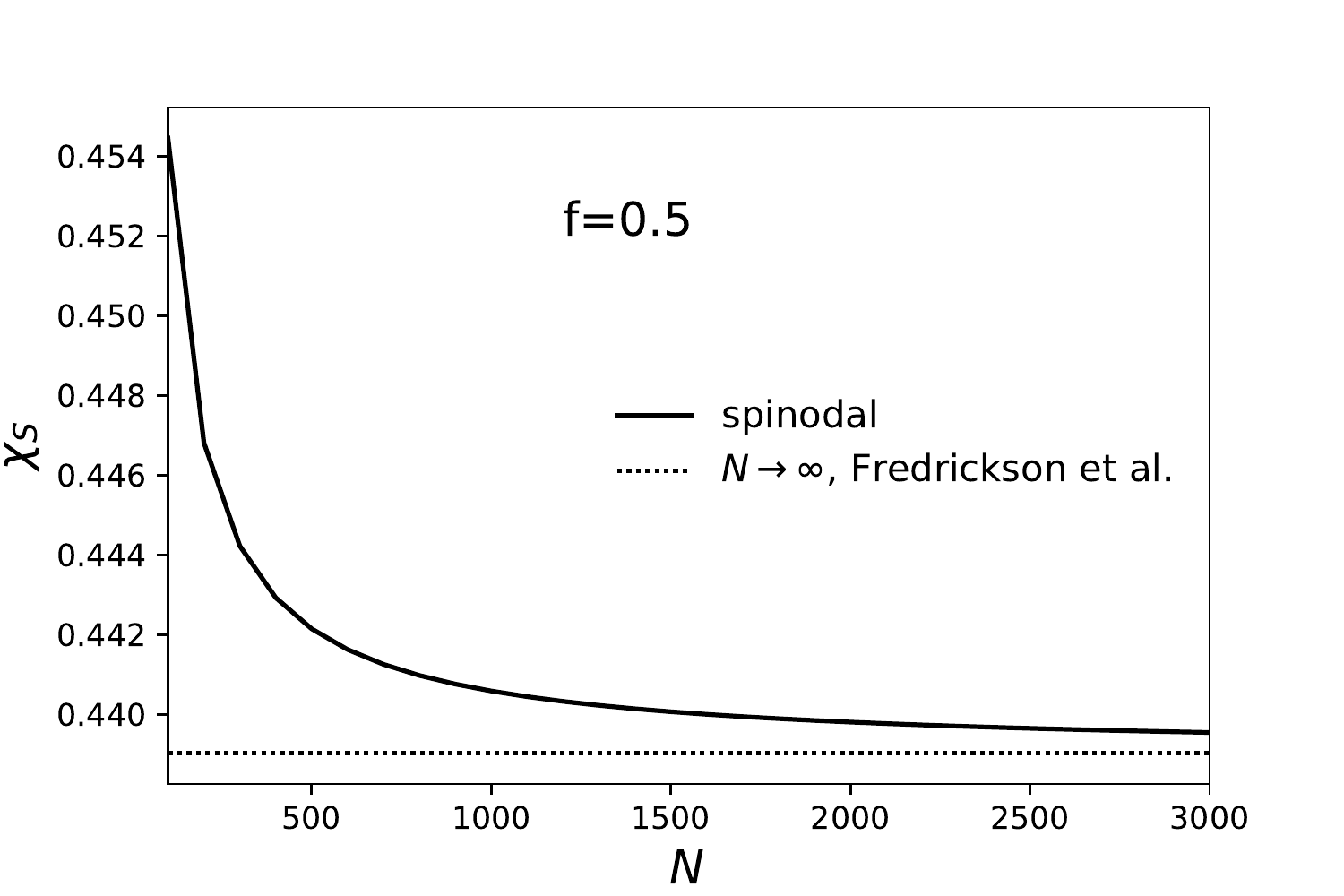}
    \caption{Spinodal points converge to the expression from the paper of Fredrickson \textit{et al.} $\chi_S=\frac{1-\lambda}{2f\left(1-f\right)\left(1+\lambda\right)}$, $f=0.5$, $\theta=0.09$.}
    \label{limit}
\end{figure}

Figure \ref{spinodal-cloud-t009} shows the dependence of cloud and spinodal points on the fraction of A-segments and chain length. As expected $\chi$ values at which spinodal and binodal points are located increase with the increasing asymmetry of the copolymer. A slight decrease of both $\chi_S$ and $\chi_C$ with $N$ can be also observed. The spinodal converges to the limit derived by Fredrickson \textit{et al.}\cite{Fredrickson1992MulticriticalMelts} for $N\rightarrow\infty$ (see Figure \ref{limit})
\begin{equation}
    {\rm{lim}}_{N\to\infty} \chi_S = \frac{\theta}{2f\left(1-f\right)\left(2f-2f^2-\theta\right)}.
\end{equation}

\begin{figure}
    \centering
    \includegraphics[width=0.7\textwidth]{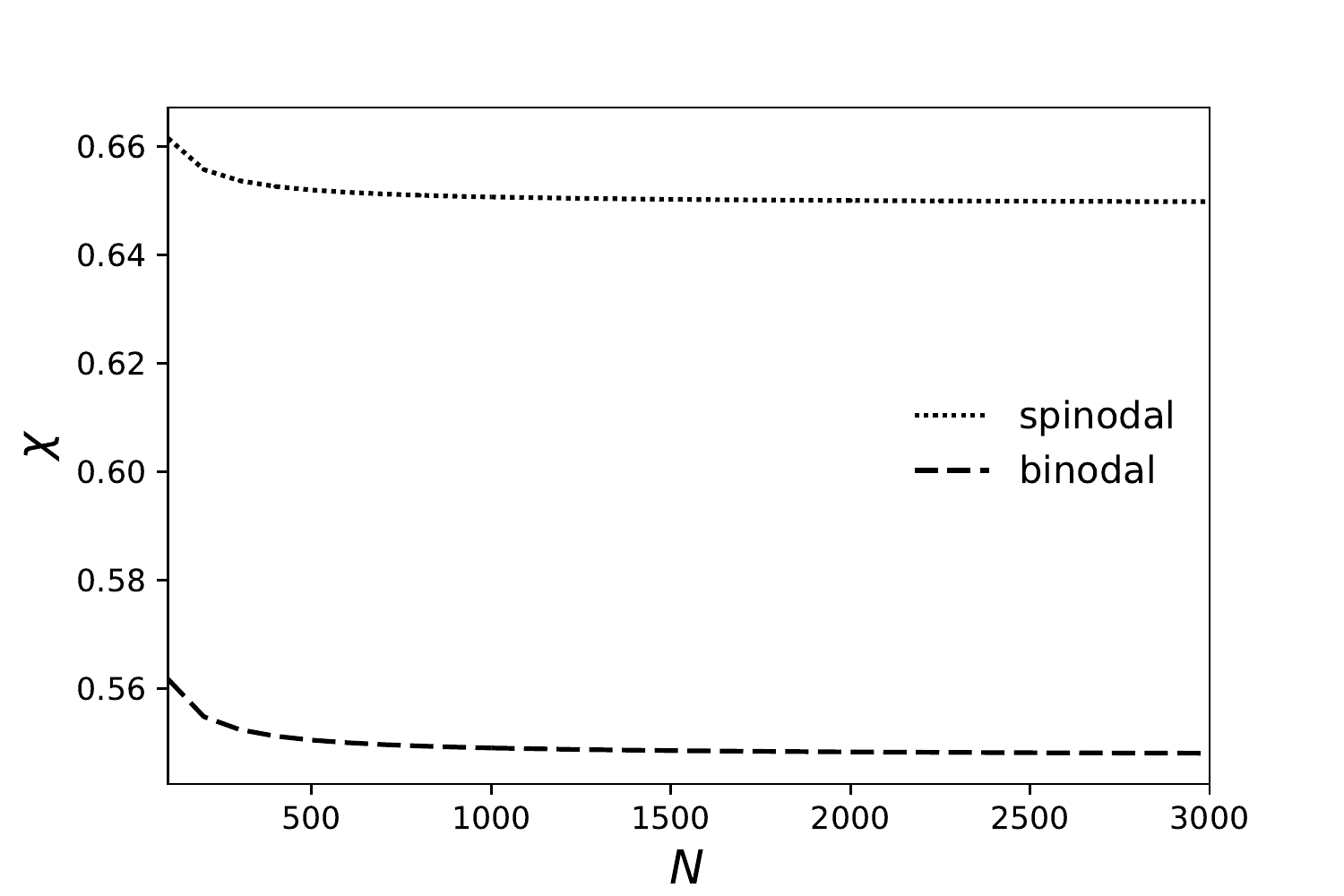}
    \caption{Dependence of spinodal points, $\chi_S$, and binodal points, $\chi_C$, on the length of the chain $N$ for copolymer with $f=0.3$, $\theta=0.09$.}
    \label{f03t009-cloud}
\end{figure}

We also predict that binodal does not converge to spinodal in the limit $N\rightarrow\infty$ (Figure \ref{f03t009-cloud}), and the difference between the Flory-Huggins parameter at the cloud point and the spinodal point, $\Delta\chi=\chi_S-\chi_C$, is nearly independent on $N$. This is a new prediction\cite{Fredrickson1992MulticriticalMelts}. 

\section{Discussion}
In the previous section, Flory-Huggins phase diagrams for correlated copolymers are presented which were calculated using the method of moments.  Here we compare obtained results with existing works and propose future developments. 

First, we want to note that our phase diagrams qualitatively look similar to those for short binomial copolymers considered by Nesarikar \textit{et al.}\cite{Nesarikar1993PhaseCopolymers}. In that paper phase diagrams were obtained by direct solution of phase equilibrium equations for all components in the system. Insofar as binomial copolymers are a special case of correlated copolymers, this demonstrates that the combination of distribution (\ref{distribution}) and the moments method produces consistent results.  

Our prediction for a spinodal converges to the expression derived by Fredrickson \textit{et. al.}\cite{Fredrickson1992MulticriticalMelts} for infinitely long chains. We also showed that the distance between binodal and spinodal is nearly independent of $N$ in contrast with previous predictions \cite{Fredrickson1992MulticriticalMelts}.    

We predict that the compositional contrast between coexisting phases is larger than the width of the distributions of these coexisting phases, so in the framework of Flory-Huggins theory phases can be always distinguished. It would be interesting to verify our predictions with respect to contrast in the composition of macrophases in simulations (at least for short sequences), for example, using the method proposed by Houdayer and M{\"u}ller \cite{Houdayer2004PhaseStudy} extended to simulate asymmetric copolymers. 

In the present work, we consider the simplest possible case of the copolymer in a melt with a fixed chain length. However, the method of moments allows taking polydispersity in chain length into account as well. The only thing which is needed for it is a distribution function depending additionally on the chain length, $\rho\left(\sigma,t,N\right)$. In the case of the PVA-PVAc system synthesized by post-modification of PVAc, this distribution function is obtained by the product of distribution function describing polydispersity of PVAc and the distribution function with respect to the fractions A-segments and AB-duplets (\ref{distribution}). 
Additionally, for the PVA-PVAc system, it should be taken into account that the volume of the VAc monomer unit is two times larger than the volume of the VA monomer unit, which leads to polydispersity in length even in the case when the initial PVAc is perfectly monodisperse. This effect is also easy to account for using the presented approach. Both polydispersity in length and segment asymmetry increase incompatibility. Interestingly, our preliminary calculations show that the segment asymmetry has a larger effect on phase diagram than the chain length polydispersity (for Poisson distribution of initial PVAc).

The proposed approach can be also generalized to predict the phase behavior of mixtures of Markov copolymers with a plasticizer (solvent) which is an important industrial problem. It is well known that the phase behavior of these mixtures is strongly affected by the polydisperse nature of these materials, especially for blocky copolymers.\cite{Sollich2007MomentSystems,Mao2018ThermodynamicAssembly, Ratzsch1991ContinuousSystems, Enders2010TheoryColumns}. 

\begin{figure}
    \centering
    \includegraphics[width=0.7\textwidth]{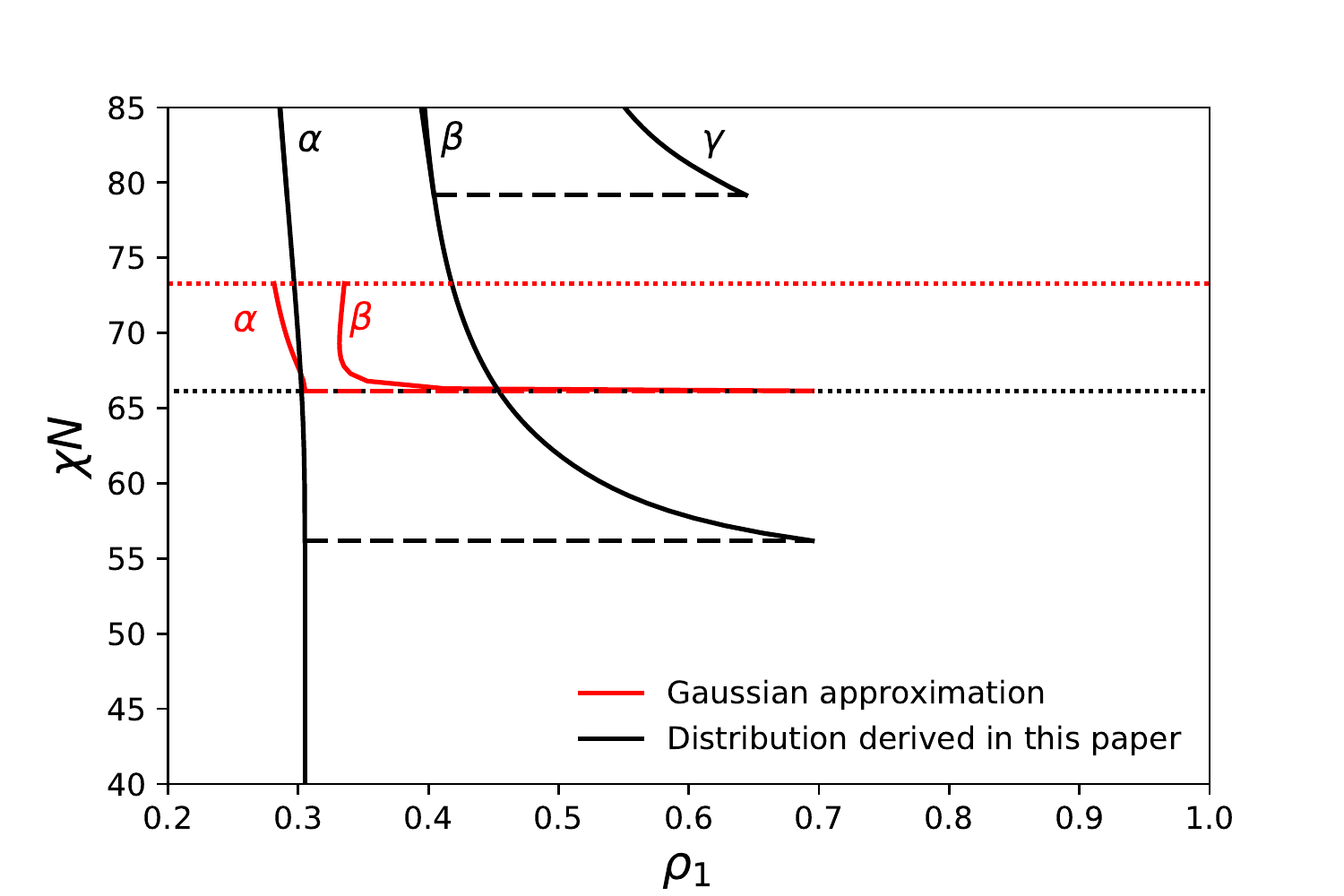}
    \caption{Comparison of composition of coexisting phases for the case of distribution (\ref{distribution}) (black) and its Gaussian approximation (red) for the case $f=0.3$, $\theta=0.09$, $N=100$. Nine moments are fixed in both cases.}
    \label{gauss}
\end{figure}

To derive a distribution function for the first-order Markov copolymers we used an approach related to the large deviation theory\cite{Touchette2009TheMechanics} which allows us to do it easily and additionally go beyond a Gaussian approximation\cite{Stockmayer1945DistributionCopolymers}. To understand the effect of large deviations in the initial distribution on phase behavior in the framework of our model we compare (see Figure \ref{gauss}) phase diagrams obtained with the method of moments for distribution (\ref{distribution}) (black curves) and the Gaussian approximation of this distribution (red curves) in a case $f=0.3$, $\theta=0.09$, $N=100$. As far as the variance for both distributions is the same, the first spinodal point is the same in both cases, it is shown by a dotted line on Figure \ref{gauss}. The composition of the shadow phase at the cloud point is also the same in both cases and equals $1-f$. The location of the first cloud point is different (shown by dashed black and red lines). In the case of Gaussian distribution, it is located just below the spinodal in contrast to the case of parent distribution (\ref{distribution}) considered in this paper. The contrast in composition between cloud ($\alpha$) and shadow ($\beta$) phases is small in the case of Gaussian distribution. It is smaller than the width of the distribution within each coexisting phase, so such coexisting phases can not be distinguished. In contrast, for the case of the distribution (\ref{distribution}) phases can be distinguished as we discussed above. This means that difference here is not just quantitative, but also qualitative. The last difference is in the location of the second spinodal point. In the case of Gaussian parent distribution, the first shadow phase loses stability at $\chi N\approx73.3$ (second spinodal point shown with a red dotted line in Figure \ref{gauss}), which is a much smaller value of $\chi N$ than in the case of parent distribution considered in this paper. We did not go beyond the second spinodal for the parent Gaussian distribution because we were unable to find a converging solution for three-phase coexistence in this case. Concluding, we can suggest that Gaussian approximation in this particular case gives qualitatively different predictions for phase behavior. 

In the present work, we use the Flory-Huggins model to describe a copolymer melt. There are several limitations to this model. One of them is that it does not allow to consider the possibility of microphase separation. It is expected that the picture including microphase separation is the following. Upon increase in Flory-Huggins parameter initially, homogeneous melt separates into two coexisting macrophases\cite{Fredrickson1992MulticriticalMelts}. At further increase in Flory-Huggins parameter, first, microphase separated phase coexisting with two macrophases is expected to emerge\cite{VonDerHeydt2011Three-phaseCopolymers}, then it is expected that macrophases will be absorbed by a microphase. The type of microphase was predicted to depend on the composition of the copolymer\cite{Angerman1996MicrophaseCopolymers,Subbotin2002PhaseCopolymers, Gavrilov2011SimulationCopolymers} and the degree of incompatibility. The possibility of coexistence between different ordered structures was also predicted\cite{Potemkin1998MicrophaseCorrections, Subbotin2002PhaseCopolymers}. All these results were obtained with the assumption that distribution in all coexisting phases had the same shape and only the average composition was different. It would be interesting to see in the future theory developments accounting both for microphase separation and fractionation for the case of asymmetric random copolymers.

\section{Conclusion}
In this paper, we derive a probability distribution function for a \textit{blocky} copolymer, with first-order Markov correlations along the sequence. Then we used this distribution with the method of moments by Sollich \textit{et al.}\cite{Sollich2007MomentSystems} and Flory-Huggins theory to obtain phase diagrams of blocky copolymers.

This combination of methods allows one to calculate Flory-Huggins phase diagrams for copolymers characterized by arbitrary length and blockiness. Qualitatively obtained phase diagrams look similar to those for short binomial copolymers\textit{et al.}\cite{Nesarikar1993PhaseCopolymers}. 

As the final message, we would like to emphasize the importance of characterizing Markov copolymers (those obtained in stationary conditions) by both fractions of A-segments and AB-duplets. This work shows that the phase behavior of two copolymers with the same composition but a different fraction of AB-duplets can be very different. So, characterization of the structure of copolymers, such as PVA, in practical settings should include the determination of this parameter.

\section{Appendix A}
In this section, we include computation details. Consider the case of moments free energy with $m$ moments fixed and two phases in equilibrium. Then to determine compositions of these phases one needs to solve a system of equations:
\begin{equation}
f\left(x_n\right)=\{\mu_{1}^{\alpha}-\mu_{_1}^{\beta},\pi^{\alpha}-\pi^{\beta},v_{\alpha}\rho_1^{\alpha}+\left(1-v_{\alpha}\right)\rho_1^{\beta}-\rho_1^{(0)},..,v_{\alpha}\rho_m^{\alpha}+\left(1-v_{\alpha}\right)\rho_m^{\beta}-\rho_m^{(0)}\}^T=0.
\end{equation}
It is solved numerically with the Newton method:
$x_{n+1}=x_{n}-f\left(x_n\right)\left[Df\left(x_n\right)\right]^{-1}$ where $x=\{\lambda_1^{\alpha},\lambda_1^{\beta},\lambda_2,...,\lambda_m,v_{\alpha}\}^T$ and:
\begin{equation}
\begin{split}
    &\left(Df\left(x_n\right)\right)_{ij}=\frac{\partial \left(f\left(x_n\right)\right)_i}{\partial \left(x_n\right)_j}=\\
    &=\begin{pmatrix}
    \frac{\partial\mu_{1}^\alpha}{\partial\lambda_1^\alpha}& -\frac{\partial\mu_{1}^\beta}{\partial\lambda_1^\beta} & \frac{\partial\left(\mu_{1}^\alpha-\mu_{1}^\beta\right)}{\partial\lambda_2} & ...& \frac{\partial\left(\mu_{1}^\alpha-\mu_{1}^\beta\right)}{\partial\lambda_m} & 0 \\
    \frac{\partial\pi^\alpha}{\partial\lambda_1^\alpha}& -\frac{\partial\pi^\beta}{\partial\lambda_1^\beta} & \frac{\partial\left(\pi^\alpha-\pi^\beta\right)}{\partial\lambda_2} & ...& \frac{\partial\left(\pi^\alpha-\pi^\beta\right)}{\partial\lambda_m} &  0  \\
    v_{\alpha}\frac{\partial\rho_1^\alpha}{\partial\lambda_1^\alpha}& \left(1-v_{\alpha}\right)\frac{\partial\rho_1^\beta}{\lambda_1^\beta} & v_{\alpha}\frac{\partial\rho_1^\alpha}{\partial\lambda_2}+\left(1-v_{\alpha}\right)\frac{\partial\rho_1^\beta}{\partial\lambda_2} & ... & v_{\alpha}\frac{\partial\rho_1^\alpha}{\partial\lambda_m}+\left(1-v_{\alpha}\right)\frac{\partial\rho_1^\beta}{\partial\lambda_m} & \rho_1^{\alpha}-\rho_1^{\beta}  \\
    ... & ... & ... & ... & ... & ... \\
    v_{\alpha}\frac{\partial\rho_m^\alpha}{\partial\lambda_m^\alpha}& \left(1-v_{\alpha}\right)\frac{\partial\rho_1^\beta}{\lambda_1^\beta} & v_{\alpha}\frac{\partial\rho_m^\alpha}{\partial\lambda_2}+\left(1-v_{\alpha}\right)\frac{\partial\rho_m^\beta}{\partial\lambda_2} & ... & v_{\alpha}\frac{\partial\rho_m^\alpha}{\partial\lambda_m}+\left(1-v_{\alpha}\right)\frac{\partial\rho_m^\beta}{\partial\lambda_m} & \rho_m^{\alpha}-\rho_m^{\beta}
    \end{pmatrix}.
    \end{split}
\label{matrix}
\end{equation}
Derivatives with respect to Lagrange parameters are calculated as:
\begin{equation}
    \frac{\partial \rho_i^{\alpha,\beta}}{\partial\lambda_j^{\alpha,\beta}}=\frac{\partial }{\partial\lambda_j^{\alpha,\beta}}\left(\frac{\int \sigma^i R_0\left(\sigma,t\right)e^{\lambda_1^{\alpha,\beta}\sigma+\sum_{k=2}^m \lambda_k \sigma^k}d \sigma d t}{\int R_0\left(\sigma,t\right)e^{\lambda_1^{\alpha,\beta}\sigma+\sum_{k=2}^m \lambda_k \sigma^k}d \sigma d t}\right)=\rho_{i+j}^{\alpha,\beta}-\rho_i^{\alpha,\beta}\rho_j^{\alpha,\beta}
\end{equation}
\begin{equation}
    \frac{\partial \lambda_0^{\alpha,\beta}}{\partial\lambda_i^{\alpha,\beta}}=\frac{\partial }{\partial\lambda_i^{\alpha,\beta}}\left(-\ln\left(\int R_0\left(\sigma,t\right)e^{\lambda_1^{\alpha,\beta}\sigma+\sum_{k=2}^m \lambda_k \sigma^k}d \sigma d t\right)\right)=-\rho_i^{\alpha,\beta}
\end{equation}

All calculations were made using scripts written in Python with the \textit{mpmath} package allowing arbitrary precision mathematics. The need in arbitrary precision mathematics arises because matrix (\ref{matrix}) in general is poorly defined.  All integrals were calculated with the trapezoidal rule and uniform discretization along the composition axis. For factorials Stirling approximation including a square root was used $n!\approx \sqrt{2 \pi n}\cdot\left(n/e\right)^n$. The step along $\chi N$ axes was varied to ensure convergence of the Newton solver and maximization of the computation speed. 

The transition from discrete distribution to continuous, as well as following discretization, introduces errors upon calculation of integrals. These errors lead to relatively large dependencies of $\lambda_i$ parameters on the degree of discretization (of order 1). These differences increase with $\chi N$ and decrease with increasing degree of discretization. Differences in $\lambda_i$, however, produce much smaller differences in free energies calculated with different discretizations, which are of order $10^{-5}$ for discretizations up to 500. However, the difference between energies of two and three phases is of order $10^{-7}$ when all calculations are done with the same discretization. As a result, the free energy of two-phase coexistence calculated with larger discretization may be smaller than the free energy of three-phase coexistence calculated with smaller discretization, even at the second spinodal point when the first shadow phase losses stability. So, it is recommended to keep discretization fixed during calculations. Values of all transition points depend to some extent on discretization. However, the difference between discretization 100 and 500 is of order $\delta \chi N \approx 0.1$ and differences tend to decrease with increase in discretization. As far as computational cost is concerned, this increases strongly with an increase of discretization ($\sim N_{discr}^{2}$). We used discretization 100 for calculations of all phase diagrams  

\section {Appendix B}
Consider again the binomial distribution:
\begin{equation}
    \rho\left(N_{\rm{A}}\right)=\frac{N!}{N_{\rm{A}}!\left(N-N_{\rm{A}}\right)!}f^{N_{\rm{A}}}\left(1-f\right)^{N-N_{\rm{A}}}
\end{equation}
In this expression $f^{N_{\rm{A}}}\left(1-f\right)^{N-N_{\rm{A}}}$ gives the probability of a specific sequence ABABB... with total length $N$ and the number of A-segments equals to $N_A$. And $\frac{N!}{N_{\rm{A}}!\left(N-N_{\rm{A}}\right)!}$ is the total number of sequences which has \textit{the same description}, i.e. have a total length $N$ and the number of A-segments equals to $N_{\rm{A}}$.
So, if we look at the distribution for a blocky copolymer
\begin{equation}
\begin{split}
    \rho\left(N_{\rm{A}},N_{\rm{AB}}\right)=&\frac{N_{\rm{A}}!\left(N-N_{\rm{A}}\right)!}{\left(N_{\rm{A}}-N_{\rm{AB}}\right)!\left(N_{\rm{AB}}!\right)^2\left(N-N_{\rm{A}}-N_{\rm{AB}}\right)!}\cdot\\
    &\cdot\frac{\left(f-\theta\right)^{N_{\rm{A}}-N_{\rm{AB}}}\theta^{2N_{\rm{AB}}}\left(1-f-\theta\right)^{N-N_{\rm{A}}-N_{\rm{AB}}}}{f^{N_{\rm{A}}}\left(1-f\right)^{N-N_{\rm{A}}}},
\end{split}
\label{distr}
\end{equation}
we can see that it has the same structure as the binomial distribution and the probability of a specific sequence, ABABB..., in which we calculated both the number of A-segments, $N_{\rm{A}}$, and the number of AB-duplets, $N_{\rm{AB}}$ equals to
\begin{equation}
p\left(ABABB...\right)=\frac{\left(f-\theta\right)^{N_{\rm{A}}-N_{AB}}\theta^{2N_{\rm{AB}}}\left(1-f-\theta\right)^{N-N_{\rm{A}}-N_{\rm{AB}}}}{f^{N_{\rm{A}}}\left(1-f\right)^{N-N_{\rm{A}}}}.
\label{prob}
\end{equation} 
So, if we calculate the probability of $ABBB$ sequence (it is short here for simplicity), we have:
\begin{equation} 
p\left(\rm{ABBB}\right)=p_{\rm{A}}\cdot p_{\rm{AB}}p_{\rm{BB}}p_{\rm{BB}}=f\cdot\frac{\theta}{f}\frac{1-f-\theta}{1-f}\cdot\frac{1-f-\theta}{1-f}=\frac{\theta\left(1-f-\theta\right)^2}{\left(1-f\right)^2}
\end{equation}
where $P_{\rm{IJ}}$ are corresponding elements of transfer matrix (\ref{matrix})
From the formula above (\ref{prob}) we obtain:
\begin{equation} 
p^*\left(\rm{ABBB}\right)=\frac{\theta^2\left(1-f-\theta\right)^2}{f\left(1-f\right)^3}
\end{equation}
So we have a discrepancy between $p\left(ABBB\right)$ and $p^*\left(ABBB\right)$. The contradiction is resolved if we take into account that in equation \ref{prob} no start of the sequence was ever specified and it must be "cyclic" without cyclic symmetry. So to obtain the probability for a linear sequence we need to specify a starting segment and remove one duplet adjacent to it (in this particular case the BA-duplet):
\begin{equation}
  p\left(\rm{ABBB}\right)=p^*\left(\rm{ABBB}\right)\cdot\frac{p_{\rm{A}}}{P_{\rm{BA}}}
\end{equation}
However, this end-chain effects will be unimportant in the limit of large $N$, $N_{\rm{A}}$ and $N_{\rm{AB}}$.

The second part of the distribution, the number of sequences with the same description $N$, $N_{\rm{A}}$ and $N_{\rm{AB}}$, can be calculated in the following way. Assuming that we have a string of A-s with length $N_{rm{A}}$ and a string of B-s with length $N-N_{\rm{A}}$. Then if we want to mix them in a way such that there are precisely $N_{\rm{BA}}+N_{\rm{AB}}=2N_{\rm{AB}}$ boundaries between A and B-blocks, we need to find a number of ways in which one can split a continuous A-string into $N_{\rm{AB}}$ blocks and for each of this splits find a number of ways the B-string can be split into $N_{\rm{AB}}$ blocks. This means that the total number of combinations is:
\begin{equation}
    \frac{N_{\rm{A}}!}{\left(N_{\rm{A}}-N_{\rm{BA}}\right)!N_{\rm{BA}}!}\cdot\frac{\left(N-N_{\rm{A}}\right)!}{\left(N-N_{\rm{A}}-N_{\rm{BA}}\right)!N_{\rm{BA}}!}
\end{equation}
This expression coincides with the number of sequences with the same description from the distribution in equation \ref{distr}. In this case, end-chain effects are also neglected as the number of boundaries $2N_{AB}$ is even, implying that all beads are located on a "circle".

It is interesting to note that distribution in equation \ref{distr} converges exactly to binomial distribution in case when $\theta=f\left(1-f\right)$ and the summation over $N_{\rm{AB}}$ is taken. This happens because for binomial distribution there is no difference between a linear chain model with ends and the circle model considered above without ends, as there is no correlation between the probability of appearance of different segments.

\begin{acknowledgement}
The authors would like to thank Procter and Gamble and the UK research council, EPSRC, for funding through grant EP/P007864/1 "Molecular Migration in Complex Matrices: Towards Predictive Design of Structured Products". 
\end{acknowledgement}


\bibliography{Mendeley}

\providecommand{\latin}[1]{#1}
\makeatletter
\providecommand{\doi}
  {\begingroup\let\do\@makeother\dospecials
  \catcode`\{=1 \catcode`\}=2 \doi@aux}
\providecommand{\doi@aux}[1]{\endgroup\texttt{#1}}
\makeatother
\providecommand*\mcitethebibliography{\thebibliography}
\csname @ifundefined\endcsname{endmcitethebibliography}
  {\let\endmcitethebibliography\endthebibliography}{}
\begin{mcitethebibliography}{42}
\providecommand*\natexlab[1]{#1}
\providecommand*\mciteSetBstSublistMode[1]{}
\providecommand*\mciteSetBstMaxWidthForm[2]{}
\providecommand*\mciteBstWouldAddEndPuncttrue
  {\def\EndOfBibitem{\unskip.}}
\providecommand*\mciteBstWouldAddEndPunctfalse
  {\let\EndOfBibitem\relax}
\providecommand*\mciteSetBstMidEndSepPunct[3]{}
\providecommand*\mciteSetBstSublistLabelBeginEnd[3]{}
\providecommand*\EndOfBibitem{}
\mciteSetBstSublistMode{f}
\mciteSetBstMaxWidthForm{subitem}{(\alph{mcitesubitemcount})}
\mciteSetBstSublistLabelBeginEnd
  {\mcitemaxwidthsubitemform\space}
  {\relax}
  {\relax}

\bibitem[Tubbs \latin{et~al.}(1966)Tubbs, I, and
  Nernours]{Tubbs1966SequencePolyvinyl-acetate}
Tubbs,~R.~K.; I,~E.; Nernours,~P.~D. {Sequence Distribution of Partially
  Hydrolyzed Poly(vinyl-acetate)}. \emph{J. Polym. Sci. Part A-1}
  \textbf{1966}, \emph{4}, 623--629\relax
\mciteBstWouldAddEndPuncttrue
\mciteSetBstMidEndSepPunct{\mcitedefaultmidpunct}
{\mcitedefaultendpunct}{\mcitedefaultseppunct}\relax
\EndOfBibitem
\bibitem[Moritani and Fujiwara(1977)Moritani, and
  Fujiwara]{Moritani197713C-Copolymers}
Moritani,~T.; Fujiwara,~Y. {13C- and 1H-NMR Investigations of Sequence
  Distribution in Vinyl Alcohol-Vinyl Acetate Copolymers}.
  \emph{Macromolecules} \textbf{1977}, \emph{10}, 532--535\relax
\mciteBstWouldAddEndPuncttrue
\mciteSetBstMidEndSepPunct{\mcitedefaultmidpunct}
{\mcitedefaultendpunct}{\mcitedefaultseppunct}\relax
\EndOfBibitem
\bibitem[Denisova \latin{et~al.}(2012)Denisova, Krentsel', Peregudov,
  Litmanovich, Podbel'skiy, Litmanovich, and
  Kudryavtsev]{Denisova2012ChainCopolymers}
Denisova,~Y.~I.; Krentsel',~L.~B.; Peregudov,~A.~S.; Litmanovich,~E.~A.;
  Podbel'skiy,~V.~V.; Litmanovich,~A.~D.; Kudryavtsev,~Y.~V. {Chain statistics
  in vinyl acetatevinyl alcohol multiblock copolymers}. \emph{Polymer Science -
  Series B} \textbf{2012}, \emph{54}, 375--382\relax
\mciteBstWouldAddEndPuncttrue
\mciteSetBstMidEndSepPunct{\mcitedefaultmidpunct}
{\mcitedefaultendpunct}{\mcitedefaultseppunct}\relax
\EndOfBibitem
\bibitem[Ilyin \latin{et~al.}(2014)Ilyin, Malkin, Kulichikhin, Denisova,
  Krentsel, Shandryuk, Litmanovich, Litmanovich, Bondarenko, and
  Kudryavtsev]{Ilyin2014EffectBulk}
Ilyin,~S.~O.; Malkin,~A.~Y.; Kulichikhin,~V.~G.; Denisova,~Y.~I.;
  Krentsel,~L.~B.; Shandryuk,~G.~A.; Litmanovich,~A.~D.; Litmanovich,~E.~A.;
  Bondarenko,~G.~N.; Kudryavtsev,~Y.~V. {Effect of chain structure on the
  rheological properties of vinyl acetate-vinyl alcohol copolymers in solution
  and bulk}. \emph{Macromolecules} \textbf{2014}, \emph{47}, 4790--4804\relax
\mciteBstWouldAddEndPuncttrue
\mciteSetBstMidEndSepPunct{\mcitedefaultmidpunct}
{\mcitedefaultendpunct}{\mcitedefaultseppunct}\relax
\EndOfBibitem
\bibitem[Denisova \latin{et~al.}(2013)Denisova, Shandryuk, Krentsel',
  Blagodatskikh, Peregudov, Litmanovich, and
  Kudryavtsev]{Denisova2013ThermalCopolymers}
Denisova,~Y.~I.; Shandryuk,~G.~A.; Krentsel',~L.~B.; Blagodatskikh,~I.~V.;
  Peregudov,~A.~S.; Litmanovich,~A.~D.; Kudryavtsev,~Y.~V. {Thermal
  fractionation of vinyl acetate-vinyl alcohol copolymers}. \emph{Polymer
  Science - Series A} \textbf{2013}, \emph{55}, 385--392\relax
\mciteBstWouldAddEndPuncttrue
\mciteSetBstMidEndSepPunct{\mcitedefaultmidpunct}
{\mcitedefaultendpunct}{\mcitedefaultseppunct}\relax
\EndOfBibitem
\bibitem[Squillace \latin{et~al.}(2020)Squillace, Fong, Shepherd, Hind, Tellam,
  Steinke, and Thompson]{Squillace2020InfluenceActivity}
Squillace,~O.; Fong,~R.; Shepherd,~O.; Hind,~J.; Tellam,~J.; Steinke,~N.~J.;
  Thompson,~R.~L. {Influence of PVAc/PVA hydrolysis on additive surface
  activity}. \emph{Polymers} \textbf{2020}, \emph{12}, 1--17\relax
\mciteBstWouldAddEndPuncttrue
\mciteSetBstMidEndSepPunct{\mcitedefaultmidpunct}
{\mcitedefaultendpunct}{\mcitedefaultseppunct}\relax
\EndOfBibitem
\bibitem[Briddick \latin{et~al.}(2018)Briddick, Fong, Sabatti{\'{e}}, Li,
  Skoda, Courchay, and Thompson]{Briddick2018BloomingFilms}
Briddick,~A.; Fong,~R.~J.; Sabatti{\'{e}},~E.~F.; Li,~P.; Skoda,~M.~W.;
  Courchay,~F.; Thompson,~R.~L. {Blooming of Smectic Surfactant/Plasticizer
  Layers on Spin-Cast Poly(vinyl alcohol) Films}. \emph{Langmuir}
  \textbf{2018}, \emph{34}, 1410--1418\relax
\mciteBstWouldAddEndPuncttrue
\mciteSetBstMidEndSepPunct{\mcitedefaultmidpunct}
{\mcitedefaultendpunct}{\mcitedefaultseppunct}\relax
\EndOfBibitem
\bibitem[Ergun \latin{et~al.}(2015)Ergun, Guo, and
  Huebner-Keese]{Ergun2015Cellulose}
Ergun,~R.; Guo,~J.; Huebner-Keese,~B. {Cellulose}. \emph{Encyclopedia of Food
  and Health} \textbf{2015}, 694--702\relax
\mciteBstWouldAddEndPuncttrue
\mciteSetBstMidEndSepPunct{\mcitedefaultmidpunct}
{\mcitedefaultendpunct}{\mcitedefaultseppunct}\relax
\EndOfBibitem
\bibitem[Karimi \latin{et~al.}(2019)Karimi, Mohammadi, and
  Hooshyari]{Karimi2019RecentReview}
Karimi,~M.~B.; Mohammadi,~F.; Hooshyari,~K. {Recent approaches to improve
  Nafion performance for fuel cell applications: A review}. \emph{International
  Journal of Hydrogen Energy} \textbf{2019}, \emph{44}, 28919--28938\relax
\mciteBstWouldAddEndPuncttrue
\mciteSetBstMidEndSepPunct{\mcitedefaultmidpunct}
{\mcitedefaultendpunct}{\mcitedefaultseppunct}\relax
\EndOfBibitem
\bibitem[Teixeira \latin{et~al.}(2007)Teixeira, Read, and
  McLeish]{Teixeira2007DemixingReactions}
Teixeira,~P.~I.; Read,~D.~J.; McLeish,~T.~C. {Demixing instability in coil-rod
  blends undergoing polycondensation reactions}. \emph{Journal of Chemical
  Physics} \textbf{2007}, \emph{126}, 074901\relax
\mciteBstWouldAddEndPuncttrue
\mciteSetBstMidEndSepPunct{\mcitedefaultmidpunct}
{\mcitedefaultendpunct}{\mcitedefaultseppunct}\relax
\EndOfBibitem
\bibitem[Daniele \latin{et~al.}(2017)Daniele, Mariconda, Guerra, Longo, and
  Giannini]{Daniele2017Single-phase14-cis-polyisoprene}
Daniele,~S.; Mariconda,~A.; Guerra,~G.; Longo,~P.; Giannini,~L. {Single-phase
  block copolymers by cross-metathesis of 1,4-cis-polybutadiene and
  1,4-cis-polyisoprene}. \emph{Polymer} \textbf{2017}, \emph{130},
  143--149\relax
\mciteBstWouldAddEndPuncttrue
\mciteSetBstMidEndSepPunct{\mcitedefaultmidpunct}
{\mcitedefaultendpunct}{\mcitedefaultseppunct}\relax
\EndOfBibitem
\bibitem[Gringolts \latin{et~al.}(2019)Gringolts, Denisova, Finkelshtein, and
  Kudryavtsev]{Gringolts2019OlefinSynthesis}
Gringolts,~M.~L.; Denisova,~Y.~I.; Finkelshtein,~E.~S.; Kudryavtsev,~Y.~V.
  {Olefin metathesis in multiblock copolymer synthesis}. \emph{Beilstein
  Journal of Organic Chemistry} \textbf{2019}, \emph{15}, 218--235\relax
\mciteBstWouldAddEndPuncttrue
\mciteSetBstMidEndSepPunct{\mcitedefaultmidpunct}
{\mcitedefaultendpunct}{\mcitedefaultseppunct}\relax
\EndOfBibitem
\bibitem[Noah \latin{et~al.}(1974)Noah, Litmanovich, and
  Plat{\'{e}}]{Noah1974ThePolymers}
Noah,~O.~V.; Litmanovich,~A.~D.; Plat{\'{e}},~N.~A. {The quantitative approach
  to the composition heterogeneity of the products of reactions of polymers}.
  \emph{Journal of Polymer Science: Polymer Physics Edition} \textbf{1974},
  \emph{12}, 1711--1725\relax
\mciteBstWouldAddEndPuncttrue
\mciteSetBstMidEndSepPunct{\mcitedefaultmidpunct}
{\mcitedefaultendpunct}{\mcitedefaultseppunct}\relax
\EndOfBibitem
\bibitem[Kim \latin{et~al.}(2020)Kim, Chakrapani, and
  Beckingham]{Kim2020TuningIsoprene}
Kim,~J.~M.; Chakrapani,~S.~B.; Beckingham,~B.~S. {Tuning Compositional Drift in
  the Anionic Copolymerization of Styrene and Isoprene}. \emph{Macromolecules}
  \textbf{2020}, \emph{53}, 3814--3821\relax
\mciteBstWouldAddEndPuncttrue
\mciteSetBstMidEndSepPunct{\mcitedefaultmidpunct}
{\mcitedefaultendpunct}{\mcitedefaultseppunct}\relax
\EndOfBibitem
\bibitem[Scott(1952)]{Scott1952ThermodynamicsCopolymers}
Scott,~R.~L. {Thermodynamics of High Polymer Solutions. VI. The Compatibility
  of Copolymers}. \emph{J. Polym. Sci.} \textbf{1952}, \emph{9}, 423--432\relax
\mciteBstWouldAddEndPuncttrue
\mciteSetBstMidEndSepPunct{\mcitedefaultmidpunct}
{\mcitedefaultendpunct}{\mcitedefaultseppunct}\relax
\EndOfBibitem
\bibitem[Bauer(1985)]{Bauer1985EquilibriumCopolymers}
Bauer,~B.~J. {Equilibrium Phase Compositions of Heterogeneous Copolymers}.
  \emph{Polym. Eng. Sci.} \textbf{1985}, \emph{25}, 1081--1087\relax
\mciteBstWouldAddEndPuncttrue
\mciteSetBstMidEndSepPunct{\mcitedefaultmidpunct}
{\mcitedefaultendpunct}{\mcitedefaultseppunct}\relax
\EndOfBibitem
\bibitem[Nesarikar \latin{et~al.}(1993)Nesarikar, Olvera De La~Cruz, and
  Crist]{Nesarikar1993PhaseCopolymers}
Nesarikar,~A.; Olvera De La~Cruz,~M.; Crist,~B. {Phase transitions in random
  copolymers}. \emph{The Journal of Chemical Physics} \textbf{1993}, \emph{98},
  7385--7397\relax
\mciteBstWouldAddEndPuncttrue
\mciteSetBstMidEndSepPunct{\mcitedefaultmidpunct}
{\mcitedefaultendpunct}{\mcitedefaultseppunct}\relax
\EndOfBibitem
\bibitem[Shakhnovich and Gutin(1989)Shakhnovich, and
  Gutin]{Shakhnovich1989FormationHeteropolymer}
Shakhnovich,~E.; Gutin,~A. {Formation of microdomains in a quenched disordered
  heteropolymer}. \emph{Journal de Physique} \textbf{1989}, \emph{50},
  1843--1850\relax
\mciteBstWouldAddEndPuncttrue
\mciteSetBstMidEndSepPunct{\mcitedefaultmidpunct}
{\mcitedefaultendpunct}{\mcitedefaultseppunct}\relax
\EndOfBibitem
\bibitem[Fredrickson \latin{et~al.}(1992)Fredrickson, Milner, and
  Leibler]{Fredrickson1992MulticriticalMelts}
Fredrickson,~G.~H.; Milner,~S.~T.; Leibler,~L. {Multicritical Phenomena and
  Microphase Ordering in Random Block Copolymers Melts}. \emph{Macromolecules}
  \textbf{1992}, \emph{25}, 6341--6354\relax
\mciteBstWouldAddEndPuncttrue
\mciteSetBstMidEndSepPunct{\mcitedefaultmidpunct}
{\mcitedefaultendpunct}{\mcitedefaultseppunct}\relax
\EndOfBibitem
\bibitem[Dobrynin and Erukhimovich(1991)Dobrynin, and
  Erukhimovich]{Dobrynin1991FluctuationSystems}
Dobrynin,~A.~V.; Erukhimovich,~I.~Y. {Fluctuation theory of weak
  crystallization in disordered heteropolymer systems}. \emph{JETP Lett.}
  \textbf{1991}, \emph{53}, 570--572\relax
\mciteBstWouldAddEndPuncttrue
\mciteSetBstMidEndSepPunct{\mcitedefaultmidpunct}
{\mcitedefaultendpunct}{\mcitedefaultseppunct}\relax
\EndOfBibitem
\bibitem[Angerman \latin{et~al.}(1996)Angerman, Brinke, and
  Erukhimovich]{Angerman1996MicrophaseCopolymers}
Angerman,~H.; Brinke,~G.; Erukhimovich,~I. {Microphase Separation in Correlated
  Random Copolymers}. \emph{Macromolecules} \textbf{1996}, \emph{29},
  3255--3262\relax
\mciteBstWouldAddEndPuncttrue
\mciteSetBstMidEndSepPunct{\mcitedefaultmidpunct}
{\mcitedefaultendpunct}{\mcitedefaultseppunct}\relax
\EndOfBibitem
\bibitem[Vanderwoude and Shi(2017)Vanderwoude, and
  Shi]{Vanderwoude2017EffectsCopolymers}
Vanderwoude,~G.; Shi,~A.~C. {Effects of Blockiness and Polydispersity on the
  Phase Behavior of Random Block Copolymers}. \emph{Macromolecular Theory and
  Simulations} \textbf{2017}, \emph{26}, 1--10\relax
\mciteBstWouldAddEndPuncttrue
\mciteSetBstMidEndSepPunct{\mcitedefaultmidpunct}
{\mcitedefaultendpunct}{\mcitedefaultseppunct}\relax
\EndOfBibitem
\bibitem[Govorun and Chertovich(2017)Govorun, and
  Chertovich]{Govorun2017MicrophaseCopolymers}
Govorun,~E.~N.; Chertovich,~A.~V. {Microphase separation in random multiblock
  copolymers}. \emph{Journal of Chemical Physics} \textbf{2017}, \emph{146},
  034903\relax
\mciteBstWouldAddEndPuncttrue
\mciteSetBstMidEndSepPunct{\mcitedefaultmidpunct}
{\mcitedefaultendpunct}{\mcitedefaultseppunct}\relax
\EndOfBibitem
\bibitem[Panyukov and Potemkin(1996)Panyukov, and
  Potemkin]{Panyukov1996TheHeteropolymers}
Panyukov,~S.~V.; Potemkin,~I.~I. {The effect of thermodynamic fluctuations on
  the formation of superstructures in random heteropolymers}. \emph{JETP Lett.}
  \textbf{1996}, \emph{64}, 197--201\relax
\mciteBstWouldAddEndPuncttrue
\mciteSetBstMidEndSepPunct{\mcitedefaultmidpunct}
{\mcitedefaultendpunct}{\mcitedefaultseppunct}\relax
\EndOfBibitem
\bibitem[Subbotin and Semenov(2002)Subbotin, and
  Semenov]{Subbotin2002PhaseCopolymers}
Subbotin,~A.; Semenov,~A. {Phase equilibria in random multiblock copolymers}.
  \emph{The European Physical Journal E} \textbf{2002}, \emph{7}, 49--64\relax
\mciteBstWouldAddEndPuncttrue
\mciteSetBstMidEndSepPunct{\mcitedefaultmidpunct}
{\mcitedefaultendpunct}{\mcitedefaultseppunct}\relax
\EndOfBibitem
\bibitem[Von Der~Heydt \latin{et~al.}(2011)Von Der~Heydt, M{\"{u}}ller, and
  Zippelius]{VonDerHeydt2011Three-phaseCopolymers}
Von Der~Heydt,~A.; M{\"{u}}ller,~M.; Zippelius,~A. {Three-phase coexistence
  with sequence partitioning in symmetric random block copolymers}.
  \emph{Physical Review E - Statistical, Nonlinear, and Soft Matter Physics}
  \textbf{2011}, \emph{83}, 1--21\relax
\mciteBstWouldAddEndPuncttrue
\mciteSetBstMidEndSepPunct{\mcitedefaultmidpunct}
{\mcitedefaultendpunct}{\mcitedefaultseppunct}\relax
\EndOfBibitem
\bibitem[Heydt \latin{et~al.}(2010)Heydt, Marcus, and
  Zippelius]{Heydt2010SequenceCopolymers}
Heydt,~A. V.~D.; Marcus,~M.; Zippelius,~A. {Sequence Fractionation in Symmetric
  Random Block Copolymers}. \emph{Macromoleculescules} \textbf{2010},
  \emph{43}, 3161--3164\relax
\mciteBstWouldAddEndPuncttrue
\mciteSetBstMidEndSepPunct{\mcitedefaultmidpunct}
{\mcitedefaultendpunct}{\mcitedefaultseppunct}\relax
\EndOfBibitem
\bibitem[Sollich \latin{et~al.}(2007)Sollich, Warren, and
  Cates]{Sollich2007MomentSystems}
Sollich,~P.; Warren,~P.~B.; Cates,~M.~E. \emph{Advances in Chemical Physics};
  John Wiley {\&} Sons, Ltd, 2007; pp 265--336\relax
\mciteBstWouldAddEndPuncttrue
\mciteSetBstMidEndSepPunct{\mcitedefaultmidpunct}
{\mcitedefaultendpunct}{\mcitedefaultseppunct}\relax
\EndOfBibitem
\bibitem[Goldie and Pinch(1991)Goldie, and
  Pinch]{Goldie1991CommunicationTheory}
Goldie,~C.~M.; Pinch,~R.~G. \emph{{Communication Theory}}; Cambridge University
  Press, 1991; p 210\relax
\mciteBstWouldAddEndPuncttrue
\mciteSetBstMidEndSepPunct{\mcitedefaultmidpunct}
{\mcitedefaultendpunct}{\mcitedefaultseppunct}\relax
\EndOfBibitem
\bibitem[Touchette(2009)]{Touchette2009TheMechanics}
Touchette,~H. {The large deviation approach to statistical mechanics}.
  \emph{Physics Reports} \textbf{2009}, \emph{478}, 1--69\relax
\mciteBstWouldAddEndPuncttrue
\mciteSetBstMidEndSepPunct{\mcitedefaultmidpunct}
{\mcitedefaultendpunct}{\mcitedefaultseppunct}\relax
\EndOfBibitem
\bibitem[Yashin \latin{et~al.}(1997)Yashin, Kudryavtsev, Govorun, and
  Litmanovich]{Yashin1997MacromolecularBlend}
Yashin,~V.; Kudryavtsev,~Y.; Govorun,~E.; Litmanovich,~A. {Macromolecular
  reaction and interdiffusion in a compatible polymer blend}.
  \emph{Macromolecular Theory and Simulations} \textbf{1997}, \emph{6},
  247--269\relax
\mciteBstWouldAddEndPuncttrue
\mciteSetBstMidEndSepPunct{\mcitedefaultmidpunct}
{\mcitedefaultendpunct}{\mcitedefaultseppunct}\relax
\EndOfBibitem
\bibitem[Kudryavtsev and Govorun(2006)Kudryavtsev, and
  Govorun]{Kudryavtsev2006Diffusion-inducedCopolymers}
Kudryavtsev,~Y.~V.; Govorun,~E.~N. {Diffusion-induced growth of compositional
  heterogeneity in polymer blends containing random copolymers}. \emph{European
  Physical Journal E} \textbf{2006}, \emph{21}, 263--276\relax
\mciteBstWouldAddEndPuncttrue
\mciteSetBstMidEndSepPunct{\mcitedefaultmidpunct}
{\mcitedefaultendpunct}{\mcitedefaultseppunct}\relax
\EndOfBibitem
\bibitem[Not()]{Note-1}
Phase behavior of stiff copolymers in the framework of wormchain model was
  recently studied by the group of A. J. Spakowitz, for example
  \cite{Mao2016ImpactCopolymers}\relax
\mciteBstWouldAddEndPuncttrue
\mciteSetBstMidEndSepPunct{\mcitedefaultmidpunct}
{\mcitedefaultendpunct}{\mcitedefaultseppunct}\relax
\EndOfBibitem
\bibitem[Balazs \latin{et~al.}(1985)Balazs, Sanchez, Epstein, Karasz, and
  MacKnight]{Balazs1985EffectBlends}
Balazs,~A.~C.; Sanchez,~I.~C.; Epstein,~I.~R.; Karasz,~F.~E.; MacKnight,~W.~J.
  {Effect of Sequence Distribution on the Miscibility of Polymer/Copolymer
  Blends}. \emph{Macromolecules} \textbf{1985}, \emph{18}, 2188--2191\relax
\mciteBstWouldAddEndPuncttrue
\mciteSetBstMidEndSepPunct{\mcitedefaultmidpunct}
{\mcitedefaultendpunct}{\mcitedefaultseppunct}\relax
\EndOfBibitem
\bibitem[Houdayer and M{\"{u}}ller(2004)Houdayer, and
  M{\"{u}}ller]{Houdayer2004PhaseStudy}
Houdayer,~J.; M{\"{u}}ller,~M. {Phase diagram of random copolymer melts: A
  computer simulation study}. \emph{Macromolecules} \textbf{2004}, \emph{37},
  4283--4295\relax
\mciteBstWouldAddEndPuncttrue
\mciteSetBstMidEndSepPunct{\mcitedefaultmidpunct}
{\mcitedefaultendpunct}{\mcitedefaultseppunct}\relax
\EndOfBibitem
\bibitem[Mao \latin{et~al.}(2018)Mao, Macpherson, Liu, and
  Spakowitz]{Mao2018ThermodynamicAssembly}
Mao,~S.; Macpherson,~Q.; Liu,~C.; Spakowitz,~A.~J. {Thermodynamic Model of
  Solvent Effects in Semiflexible Diblock and Random Copolymer Assembly}.
  \emph{Macromolecules} \textbf{2018}, \emph{51}, 4167--4177\relax
\mciteBstWouldAddEndPuncttrue
\mciteSetBstMidEndSepPunct{\mcitedefaultmidpunct}
{\mcitedefaultendpunct}{\mcitedefaultseppunct}\relax
\EndOfBibitem
\bibitem[R{\"{a}}tzsch and Wohlfarth(1991)R{\"{a}}tzsch, and
  Wohlfarth]{Ratzsch1991ContinuousSystems}
R{\"{a}}tzsch,~M.~T.; Wohlfarth,~C. {Continuous thermodynamics of copolymer
  systems}. \emph{Advances in Polymer Science} \textbf{1991}, \emph{98},
  48--114\relax
\mciteBstWouldAddEndPuncttrue
\mciteSetBstMidEndSepPunct{\mcitedefaultmidpunct}
{\mcitedefaultendpunct}{\mcitedefaultseppunct}\relax
\EndOfBibitem
\bibitem[Enders(2010)]{Enders2010TheoryColumns}
Enders,~S. In \emph{Polymer Thermodynamics: Liquid Polymer-Containing
  Mixtures}; Wolf,~B.~A., Enders,~S., Eds.; Springer Berlin Heidelberg, 2010;
  pp 271--328\relax
\mciteBstWouldAddEndPuncttrue
\mciteSetBstMidEndSepPunct{\mcitedefaultmidpunct}
{\mcitedefaultendpunct}{\mcitedefaultseppunct}\relax
\EndOfBibitem
\bibitem[Stockmayer(1945)]{Stockmayer1945DistributionCopolymers}
Stockmayer,~W.~H. {Distribution of chain lengths and compositions in
  copolymers}. \emph{The Journal of Chemical Physics} \textbf{1945}, \emph{13},
  199--207\relax
\mciteBstWouldAddEndPuncttrue
\mciteSetBstMidEndSepPunct{\mcitedefaultmidpunct}
{\mcitedefaultendpunct}{\mcitedefaultseppunct}\relax
\EndOfBibitem
\bibitem[Gavrilov \latin{et~al.}(2011)Gavrilov, Kudryavtsev, Khalatur, and
  Chertovich]{Gavrilov2011SimulationCopolymers}
Gavrilov,~A.~A.; Kudryavtsev,~Y.~V.; Khalatur,~P.~G.; Chertovich,~A.~V.
  {Simulation of phase separation in melts of regular and random multiblock
  copolymers}. \emph{Polymer Science - Series A} \textbf{2011}, \emph{53},
  827--836\relax
\mciteBstWouldAddEndPuncttrue
\mciteSetBstMidEndSepPunct{\mcitedefaultmidpunct}
{\mcitedefaultendpunct}{\mcitedefaultseppunct}\relax
\EndOfBibitem
\bibitem[Potemkin and Panyukov(1998)Potemkin, and
  Panyukov]{Potemkin1998MicrophaseCorrections}
Potemkin,~I.~I.; Panyukov,~S.~V. {Microphase separation in correlated random
  copolymers: Mean-field theory and fluctuation corrections}. \emph{Physical
  Review E - Statistical Physics, Plasmas, Fluids, and Related
  Interdisciplinary Topics} \textbf{1998}, \emph{57}, 6902--6912\relax
\mciteBstWouldAddEndPuncttrue
\mciteSetBstMidEndSepPunct{\mcitedefaultmidpunct}
{\mcitedefaultendpunct}{\mcitedefaultseppunct}\relax
\EndOfBibitem
\end{mcitethebibliography}

\end{document}